\documentclass[prb,aps,twocolumn,superscriptaddress,preprintnumbers,nopacs, longbibliography]{revtex4-1}

\usepackage{times}
\usepackage{amsmath}
\usepackage{amssymb}
\usepackage{amsbsy}
\usepackage{amsfonts}
\usepackage{graphicx}
\usepackage{hyperref}
\usepackage{amsthm}
\usepackage{framed}
\usepackage{comment}
\usepackage{resizegather}
\usepackage{xr}

\newcommand{\bfig}{\begin{center}\vskip 0.5em}
\newcommand{\efig}{\end{center}\vskip 0.5em}
\newcommand{\eqnref}[1]{Eq.~(\ref{#1})}
\newtheorem{thm}{Theorem}

\newcommand{\ket}[1]{|#1\rangle}
\newcommand{\bra}[1]{\langle#1|}

\usepackage{color}

\newcommand{\de}[1]{[\emph{\color{green}{DE: #1}}]}
\newcommand{\jk}[1]{[\emph{\color{cyan}{JK: #1}}]}
\newcommand{\Tr}{\mathrm{Tr}}

\begin{document}


\title{Prethermal Strong Zero Modes and Topological Qubits}

\author{Dominic V. Else}
\affiliation{Physics Department, University of California,  Santa Barbara, CA 93106, USA}

\author{Paul Fendley}
\affiliation{All Souls College, University of Oxford, OX1 4AL, UK}
\affiliation{Rudolf Peierls Centre for Theoretical Physics, 1 Keble Road, Oxford, OX1 3NP, UK}

\author{Jack Kemp}
\affiliation{Rudolf Peierls Centre for Theoretical Physics, 1 Keble Road, Oxford, OX1 3NP, UK}

\author{Chetan Nayak}
\affiliation{Station Q, Microsoft Research, Santa Barbara, CA 93106, USA}
\affiliation{Physics Department, University of California,  Santa Barbara, CA 93106, USA}

\begin{abstract}
We prove that quantum information encoded in some topological excitations,
including certain Majorana zero modes,
is protected in closed systems
for a time scale exponentially long in system parameters. This protection holds even at infinite temperature. At lower temperatures the decay time becomes even longer, with a temperature dependence controlled by an effective gap that is parametrically
larger than the actual energy gap of the system. This non-equilibrium dynamical phenomenon is a form of
prethermalization, and occurs because of obstructions to the equilibriation of edge
or defect degrees of freedom with the bulk.
We analyze the ramifications for ordered and topological phases in 
one, two, and three dimensions, with examples including Majorana and
parafermionic zero modes in interacting spin chains. Our results are based on a non-perturbative analysis valid in any dimension, and they are illustrated by numerical simulations in one dimension.
We discuss the implications for experiments on quantum-dot chains tuned into a regime
supporting end Majorana zero modes, and on trapped ion chains.
\end{abstract}

\maketitle

\section{Introduction}
Solid-state systems supporting non-Abelian anyons, such as Majorana zero modes (MZMs),
are the focus of considerable research aimed at exploiting them for
quantum information processing \cite{Nayak08,Alicea12a,Beenakker13a,DasSarma15}.
In the limit of zero temperature, the quantum information stored in a collection of non-Abelian anyons
is protected, up to corrections exponentially small in the separation between anyons. At non-zero
temperatures, however, thermally excited
bulk quasiparticles can be absorbed or emitted by a zero mode, thereby corrupting the quantum information
contained therein. Thus a separation-independent failure of protection is expected to
increase with temperature as $e^{-\Delta/T}$, where $\Delta$ is an energy gap. These processes increase the width and reduce the height of the predicted \cite{Sau10a,Lutchyn10,Oreg10} zero-bias peak that appears to have been observed in tunneling experiments \cite{Mourik12,Rokhinson12,Deng12,Churchill13,Das12,Finck12,Albrecht16}.

However, the decay of a zero mode and the quantum information encoded in it is a non-equilibrium
dynamical process, and it is not clear if thermodynamic reasoning can describe it properly. In the absence of electron-electron interactions -- for instance, in the Kitaev chain
Hamiltonian \cite{Kitaev01} or the transverse-field Ising chain, to which it is related by a Jordan-Wigner
transformation -- thermally excited quasiparticles do not affect the zero modes at all. This can be reconciled
with the previous paragraph by noting that, in the absence of interactions, the coefficient in front of $e^{-\Delta/T}$ vanishes. While the absence of interactions is a fine-tuned
special case, a similar conclusion holds in systems with strong disorder in which many-body
localization\footnote{For more on many-body localization, see Ref. \onlinecite{Nandkishore14}
and references therein.} occurs. Here disorder-induced localization prevents bulk excitations from
carrying quantum information away from a zero mode \cite{Huse13,Bauer13,Bahri15}. 

Disorder, however, is not necessary to have zero modes in interacting systems. In at least one integrable system, the XYZ spin chain, exact edge zero modes survive the presence of interactions \cite{Fendley16}. This edge ``strong zero mode'' is an operator that commutes with the Hamiltonian up to exponentially small corrections in the finite size of the system  \cite{Fendley12,Jermyn14,Alicea15}. Moreover, as with the transverse-field Ising chain, thermally excited quasiparticles do not cause the edge degrees of freedom to equilibriate with the bulk. Rather, the edge spin coherence lasts forever in a semi-infinite chain, even at infinite temperature  \cite{Kemp17}. Even more strikingly, similar behavior was found in several non-integrable deformations of the Ising chain. Here the coherence time is not infinite, but extremely long-lived \cite{Kemp17}.

The purpose of this paper is to show that such long-lived edge modes are a more general phenomenon,
and to give a direct and rigorous method for understanding them. We demonstrate that ``prethermalization,''
the exponentially slow approach to thermal equilibrium that occurs in some closed quantum systems \cite{Berges04,Moeckel08,Kollar11,Gring12,Essler14,Abanin15b}, can protect edge zero modes
and, in fact, topological degrees of freedom in higher-dimensional systems as well.
(Prethermalization can also occur in periodically driven systems
\cite{Bukov15,Abanin15a,Abanin15b,Canovi16,Abanin17,Kuwahara15,Mori16,Bukov16,Else17a},
but this is not our focus here.)  Our analysis gives a clear meaning to the notion of
an ``almost'' strong zero mode: it is an operator that commutes with the full Hamiltonian
of a system up to corrections that are a nearly exponentially small function of a ratio of energy scales. Here by ``nearly'' exponentially small we mean with a logarithmic correction to the exponent, as set out in equation \eqref{nstardef} below.
We call such an operator a ``prethermal strong zero mode''.
Its lifetime is bounded below by a nearly exponentially growing
function of this ratio of energy scales because prethermalization
delays equilibration of a prethermal strong zero mode until this late time.

By relating the protection of quantum information to prethermalization,
we reveal the limits of such protection. We elucidate the nature of this protection
in one-dimensional, two-dimensional, and three-dimensional closed systems.
However, solid-state devices are not closed systems and prethermalization in such
devices is eventually superseded by thermalization driven by electron-phonon interactions.
Thus, the applicability of these ideas to Majorana zero modes in semiconductor-superconductor devices
depends on the particular device considered (since the prethermal limit is not accessible
in all devices) and, even then, will only be in some temperature range over which
the electron-phonon interaction does not dominate. 
We quantitatively analyze a quantum dot chain proposed in Refs. \onlinecite{Sau12,Fulga13};
although it has not been realized in experiments yet, it is a useful case study.
We show that prethermalization can occur
over a range of time scales.
In this prethermal regime, arguments relying on thermal equilibrium
are ultimately correct, but with a (nearly) exponentially small prefactor that reflects the slow thermalization
of the system. Moreover, the naive energy gap $\Delta$ is replaced with a much larger
effective energy gap $\Delta_\text{eff}$. This suggests that the $T>0$ protection of quantum information
may be optimized by entering the prethermal regime, in addition to -- or even rather than -- maximizing
the energy gap. We demonstrate this tradeoff in explicit models.

The type of prethermalization we describe is not special
to one-dimensional or topological systems. We describe explicitly how analogous phenomena occur in some two- and three-dimensional systems, and how prethermalization protects edge modes for long times in systems not topologically ordered. One particular example we describe in detail is the transverse-field Ising chain perturbed by integrability-breaking interactions. While this chain is related to the quantum-dot chain via a Jordan-Wigner transformation, the non-locality of the map means that topological order in the latter is simply ordinary ferromagnetic order in the former. Nonetheless, we show how prethermalization means the edge spin coherence lasts for very long times here as well under {\em any} perturbation preserving the ${\mathbb Z}_2$ spin-flip symmetry. Namely, the spin coherence lasts for a time that is (nearly) exponentially large in terms of the couplings.

We begin with a conceptual overview in Section \ref{sec:conceptual}.
In section~\ref{sec:prethermal}, we explain the prethermal regime and the theorem due to Abanin \textit{et al.}\cite{Abanin15b} that guarantees its existence for certain Hamiltonians. In the first part of Section~\ref{sec:prethermal-protection}, we show how this theorem can be used to provide a lower bound on the lifetime of edge zero modes, and apply it explicitly to a model of interacting Majorana fermions. In the remainder of Section~\ref{sec:prethermal-protection}, we discuss the lifetime of the Majorana zero modes at non-zero temperatures and present numerical simulations supporting our arguments. We then generalize this analysis in the following Section \ref{sec:general}, describing the conditions needed to observe prethermally protected zero modes, and giving details of several examples which illustrate these conditions. In section~\ref{sec:highd}, we apply this general strategy to analyze systems in two and three dimensions.
Section~\ref{sec:experiment} explores possible practical applications of our results, in quantum dot and ion chains.
Finally, in Section~\ref{sec:integrable}, we consider integrable systems, where the zero modes may survive much longer (possibly even infinitely longer)
than the lower bound.


\section{Topological zero modes at finite temperature}
\label{sec:conceptual}

A common feature of systems exhibiting topological order at zero temperature is topological degeneracy, where there are several nearly degenerate ground states. Their energy splitting scales as $e^{-L/\xi}$ for some correlation length $\xi$, where $L$ is the system size. Moreover, these degenerate states are locally indistinguishable.
This means that, at zero temperature, quantum information can be stored in the degenerate ground state subspace in a topologically protected way. However, this topological protection usually does not extend to finite temperature. Here we will review, in a schematic way, the standard arguments for this. We then explain why, in an \emph{isolated} system, prethermalization can improve the situation considerably, both in topologically and conventionally ordered systems. A rigorous argument will be given in later sections.

\begin{figure}
\includegraphics[scale=1.5]{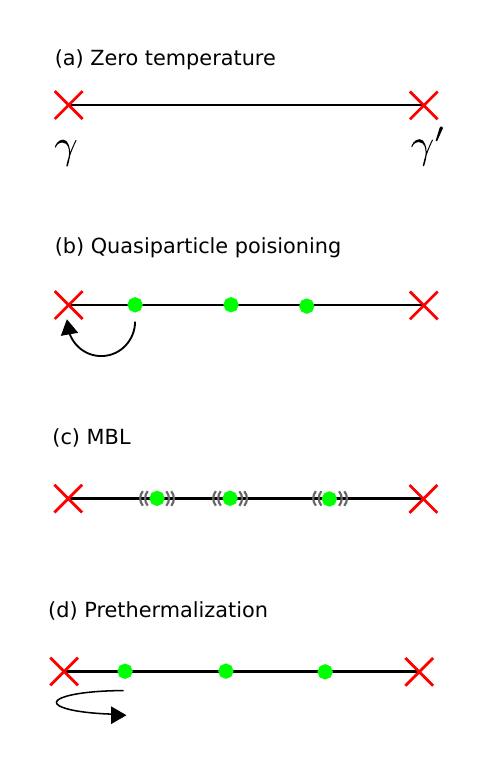}
\caption{(a) A 1-D chain of topological superconductor, with Majorana zero modes at the edges. (b) At finite temperature, mobile quasiparticles can annihilate on the Majorana zero modes, decohering the quantum information stored. (c) With strong disorder, the chain can be made to be MBL, such that the localized quasiparticles are not able to annihilate on the boundaries. (d) In a suitable ``pre-thermal'' regime, the quasiparticles are mobile but are prevented from annihilating on the boundary by an approximate conservation law.\label{conceptual_figure}}
\end{figure}

Our prototypical example is a one-dimensional topological superconductor, exemplified by the Kitaev chain \cite{Kitaev01}. In such systems with open boundary conditions, there is a pair of Majorana zero modes on the two ends of the chain, represented by Majorana operators $\gamma$ and $\gamma$' [see Figure \ref{conceptual_figure}(a)]. At zero temperature, these can be used to encode quantum information in the qubit. The qubit can be decohered by randomly acting on it with the logical operators $\sigma^z = i\gamma \gamma'$ or $\sigma^x = \gamma$. Since these are both non-local (the latter because it is a fermionic operator), the qubit is therefore immune to decoherence from any local noise process.

At finite temperature, on the other hand, there is a finite density of fermionic quasiparticles in the bulk of the system. If such a quasiparticle is near one of the Majorana zero modes (say the one corresponding to $\gamma$), then it can annihilate on it [Figure \ref{conceptual_figure}(b)], which has the effect of acting on the encoded qubit with the logical operator $\sigma^x = \gamma$. At finite temperature, such processes will happen continuously, and so the encoded qubit will quickly decohere. This is known as ``quasiparticle poisoning''.

How can quasiparticle poisoning be overcome? One way is by many-body
localization (MBL)
\cite{Basko06a,Basko06b,Oganesyan07,Pal10,Bardarson12,Serbyn13a,Serbyn13b,Bauer13,Imbrie14,Huse14}.
This causes the quasi-particles to become immobile, and thus prevents them from
moving onto the boundary and annihilating \cite{Huse14}, as shown in Figure
\ref{conceptual_figure}(c). Going to an MBL phase, however, requires strong
disorder, and it is not at all clear if such phases exist outside one dimension.
Moreover, MBL systems may not be suitable for topological quantum computation with
Majorana zero modes due to non-local rearrangements that occur when varying the
Hamiltonian adiabatically \cite{Khemani15a,Johri16}.

We show in this work that there is another way of avoiding quasiparticle poisoning. Our approach exploits the fact that if the number of quasiparticles in the bulk is conserved (in fact, it is sufficient for the number to be conserved modulo 2), then they cannot be annihilated on the boundaries [Figure \ref{conceptual_figure}(d)]. Of course, outside integrable systems there is no reason why such a conservation law should hold exactly, but often parameters of the Hamiltonian can be tuned such that it holds in an approximate way, leading to a long decoherence time.

One might imagine that obtaining a long decoherence time in this way would require significant fine-tuning. Remarkably, this turns out not to be the case. We show that many Hamiltonians possess a significant parameter regime with an approximately conserved quantity that we identify with quasiparticle number. Terms violating the conservation law are \emph{exponentially} small in parameters of the Hamiltonian. This is based on the mechanism of prethermalization, as discussed further in later sections.

Our approach is not necessarily limited to one-dimensional systems. In higher dimensions, decoherence of quantum information is also generally related to processes involving quasi-particles. For example, in the 2D toric code defined on a torus, decoherence is caused by quasiparticles moving around a non-contractible loop on the torus. On the other hand, the 4D toric code is known to be immune from decoherence at low temperatures, precisely because this system has no topologically non-trivial particle-like excitations \cite{Dennis02}.  Unfortunately, no such system is known in dimension less than four.

We show that prethermalization can be used to suppress decoherence in an isolated system arising from creation or annihilation of quasiparticles. Thus it is no help in the case of the toric code on a torus, since quasiparticles can move around non-contractible loops without changing the total quasiparticle number. However, in the 
\emph{planar }version of the toric code the decoherence mechanism involves quasiparticles annihilating on the edges.  Prethermalization therefore is useful here. We analyze this and other higher-dimensional examples in more detail in Section \ref{sec:highd}.

\section{Prethermal Regime}
\label{sec:prethermal}
A closed quantum system is said to be ``prethermal'' if, en route to thermalization,
it is in an exponentially long-lived quasi-steady state. One cause of
prethermalization is an approximate conservation law: over intermediate time scales -- known as the prethermal
regime -- the system maximizes its entropy,
subject to the constraint that the conserved quantity
takes a fixed value. Over sufficiently long time scales the entropy is maximized without any constraint.

A theorem due to Abanin, De Roeck, Huveneers and Ho (henceforth ADHH) \cite{Abanin15b} guarantees the existence of such a prethermal regime
for Hamiltonians of the form
\begin{equation}
\label{eqn:basic-Hamiltonian}
\hat{H} = -J \hat{N} + \hat{Y},
\end{equation}
where  $\hat{N}$ is a sum of finite-range commuting terms, such that $\hat{N}$ has integer eigenvalues, i.e.\ $e^{2\pi i \hat{N}} = 1$. The proof in Ref.\ \onlinecite{Abanin15b} assumes that each term in $\hat{N}$ acts only on a single site, but in Appendix \ref{app:nonsinglesite} we show that this assumption can be relaxed.
We define a parameter $J_0$
that is essentially the largest operator norm of any local
term in $Y$; more explicit forms are given in specific examples below.
The theorem says that for $J_0/J$ sufficiently small, there exists a local unitary transformation $\mathcal{U}$ such that
\begin{equation}
\label{eqn:rotated-H=N+D+E}
\mathcal{U} \hat{H} \mathcal{U}^{\dagger} = -J \hat{N} + \hat{D} + \hat{E}
\end{equation}
where $[\hat{N},\hat{D}]=0$ and $\| \hat{E} \| = O(e^{-c n_*})$ where
\begin{equation}
n_* = \left\lceil \frac{J/J_0}{[1+\log(J/J_0)]^3} \right \rceil
\label{nstardef}
\end{equation}
and $c$ is a constant. Since we focus on the $J\gg J_0$ limit, we often drop the $1$ in the
denominator of Eq. (\ref{nstardef}) for the sake of uncluttering the equations.
As a result, the dynamics of the system conserves $\hat{N}$ until a time $t_* \propto e^{c n_*}$.
A more precise statement of the ADHH theorem can be found in Ref. \onlinecite{Abanin15b}.

For our purposes, the essential point of the ADHH theorem is that a Hamiltonian (\ref{eqn:basic-Hamiltonian})
has an emergent approximate $U(1)$ symmetry generated by $\hat{N}$. The symmetry violations come from terms in the transformed Hamiltonian that are nearly \emph{exponentially small} in the large-$J$ limit.
We apply the ADHH theorem to Hamiltonians with edge zero modes, and show that the approximate $U(1)$ symmetry can protect these
zero modes -- even far from the ground state of the system, where we do not
ordinarily expect topological protection.  It also protects them in systems with no topological order whatsoever, thus providing a completely different mechanism for preserving quantum coherence distinct from topological considerations.

There is a nice heuristic picture for why a Hamiltonian of the form (\ref{eqn:basic-Hamiltonian}) has an approximate $U(1)$ symmetry. The transformation to the form (\ref{eqn:rotated-H=N+D+E}) means that it is very difficult for the terms in $\hat{Y}$ to
cause transitions between different eigenspaces of $\hat{N}$: in order
to conserve energy, many excitations of $\hat{Y}$ must be created or annihilated.
Such a process occurs slowly, so violation of the approximate $U(1)$ symmetry takes a very long time.

However, in applying these ideas to real systems, it is important to keep in mind that the ADHH
theorem requires that $\hat{Y}$ be a sum of local terms, each of which has bounded (indeed, small) norm.
This condition cannot be satisfied in any real solid since phonons and photons do not have
finite-dimensional local Hilbert spaces. The energy
associated with a transition between different eigenspaces of $\hat{N}$ need not
wait for many excitations of $\hat{Y}$ to be created or annihilated; it can, instead,
be supplied or carried away by a phonon or photon.
However, if the couplings of the electronic degrees of freedom to phonons and photons
are sufficiently small, they may not play a role during the prethermal regime. We show that this
is the case in semiconductor devices in Section \ref{sec:quantum-dot-chains}. We note also that the presence of gapless excitations does not automatically destroy all symmetry-protected edge modes \cite{Scaffidi17}.

\section{Prethermal protection of a topological qubit}
\label{sec:prethermal-protection}

The importance of the ADHH theorem for the present work is that in many cases $\mathcal{U} \hat{N} \mathcal{U}^{\dagger}$ can be identified with an effective ``quasi-particle number'', and its approximate conservation can suppress the decoherence mechanisms of a topological qubit, as outlined in section \ref{sec:conceptual}. 

In this section we introduce the idea by describing a particular example in depth:
a topological superconducting chain. We show that it is possible to construct a
prethermal edge Majorana zero mode for arbitrary interactions,
as long as the topological superconducting ordering term is the dominant coupling.
We present evidence from numerical simulations that the prethermal regime persists over
a surprisingly large range of couplings, including values of the dominant coupling that are not so very large.
Later, in section \ref{sec:quantum-dot-chains}, we discuss the relevance of
these ideas to quantum dot chains in a semiconductor-superconductor heterostructure, where the electron-phonon coupling cuts off prethermalization.
As we will see, there are circumstances under which the effect of the electron-phonon coupling is weaker than
electron-electron interactions, so that prethermalization acts to suppress the dominant MZM decay channel,
leading to relatively long-lived MZMs.
Although this first example is one-dimensional, the basic idea and
the theorem on which it relies work in any dimension, as we discuss in later sections.  

\subsection{Prethermalization in the interacting Kitaev chain}
\label{sec:interacting_majorana}

The Kitaev chain \cite{Kitaev01} is a simplified model of a one-dimensional topological superconductor
of spinless fermions. Its Hamiltonian is
\begin{align}
\nonumber
H = -\sum_j \Big[t\, &(c^\dagger_{j} c^{}_{j+1} + c^{}_{j} c^{\dagger}_{j+1})+
\mu\, c^\dagger_{j} c^{}_{j}\\
 & +\ \Delta (c^{}_{j} c^{}_{j+1} + c^{\dagger}_{j+1} c^{\dagger}_{j}) \Big]\ .
 \label{eqn:H-Kitaev}
\end{align}
The topological superconducting phase occurs for $2|t|>|\mu|$ provided $\Delta>0$.

Moreover, we will need to include interactions in our description.
They can be and usually are dropped for the purposes of demonstrating the existence
of the topological superconducting phase and its concomitant MZMs,
but they are necessary for any discussion of non-zero temperature dynamics.
With the addition of such terms, Eq. (\ref{eqn:H-Kitaev}) takes the form
\begin{multline}
\label{eqn:H-Kitaev+int}
H = \sum_j \Big[-\frac{t+\Delta}{2} \left(c^\dagger_{j} c^{}_{j+1} + c^{}_{j} c^{\dagger}_{j+1}+
c^{}_{j} c^{}_{j+1} + c^{\dagger}_{j+1} c^{\dagger}_{j}\right)\\
-\frac{t-\Delta}{2} \left(c^\dagger_{j} c^{}_{j+1} + c^{}_{j} c^{\dagger}_{j+1} -
c^{}_{j} c^{}_{j+1} - c^{\dagger}_{j+1} c^{\dagger}_{j} \right)\\
 - \,\, \mu c^\dagger_{j} c^{}_{j} \,\, +\, \, V c^\dagger_{j} c^{}_{j} c^\dagger_{j+1} c^{}_{j+1}
+ \ldots \Big]
\end{multline}
We have only written the simplest interaction term explicitly (the $V$ term) and denoted the rest implicitly with the ellipses.
In writing the Hamiltonian in this way, we have explicitly separated the presumed largest term, written on the first line,
from the smaller terms, written on the second and third lines.

This model can be written in terms of the Majorana fermion operators $\gamma_j^A$ and $\gamma_J^B$
defined by
$c_j = (\gamma_j^A + i \gamma_J^B)/2$. This rewriting gives $\hat{H}=-J\hat{N} + \hat{Y}$, where
\begin{equation}
\label{eqn:MZM-L}
\hat{N} = i\sum_{j=1}^{L-1} \gamma_j^B \gamma_{j+1}^A = \sum_{j=1}^{L-1} \sigma^z_i \sigma^z_{i+1}\ ,
\end{equation}
and $J \equiv (t+\Delta)/2$.
On the right-hand side of the second equal sign, we have written the Jordan-Wigner transformed
representation of this operator according to the definitions
\begin{align}
\gamma_{j}^A\equiv \sigma^z_j \prod_{k=1}^{j-1}\sigma^x_k\ ,\quad \gamma_j^B=i\sigma^x_j\gamma_j^A\ .
\label{JWdef}
\end{align}
Note that we have taken open boundary conditions,
in order to focus on the physics of the edge. One important thing to note is that with open boundary conditions, the two Majorana fermions at the edges, $\gamma_1^A$ and $\gamma_L^B$, do not appear in $\hat{N}$.
Each therefore commutes with it. We also have
\begin{align}
\nonumber
\hat{Y} = &-ih \sum_{j=1}^L \gamma_j^A \gamma_{j}^B
- {\sum_{\alpha, \beta}}  J_{\alpha\beta}\sum_{j=1}^{L-1} i\gamma_j^\alpha \gamma_{j+1}^\beta\\
\nonumber
&- h_2 \sum_{j=1}^{L-1} \gamma_j^A \gamma_{j}^B \gamma_{j+1}^A \gamma_{j+1}^B
- J_2 \sum_{j=1}^{L-2} \gamma_j^B  \gamma_{j+1}^A \gamma_{j+1}^B \gamma_{j+2}^A\\
&+\ldots
\label{eqn:MZM-Y}
\end{align}
where $\ldots$ denotes other third-neighbor and more-distant
hopping and interaction terms. 
The transverse field $h$ is the chemical potential of the topological superconductor (\ref{eqn:H-Kitaev+int})
according to the identification $h \equiv \mu$.
The two four-Fermi terms we single out are the two simplest, and in spin language are ${h_2}\sigma^x_j \sigma^x_{j+1}$ and $J_2\sigma^z_j\sigma^z_{j+2}$ respectively; in terms of the original topological superconductor,
$h_2 \equiv V$ while $J_2$ includes slightly longer-ranged interactions and Cooper pair-hopping terms not
explicitly included in Eq. (\ref{eqn:H-Kitaev+int}).
For simplicity, we take couplings in $\hat{Y}$ to be spatially uniform, but this is not necessary for the approach to work; in fact adding disorder to $\hat{Y}$ typically enhances the effects that we describe.
We thus have included a $J_{AB}$ term in Eq. (\ref{eqn:MZM-Y}); because of the integer-eigenvalue restriction it cannot 
be absorbed into the $J\hat{N}$ term if it is disordered.
We assume that the hopping and interaction terms have finite range but make no further assumptions.


We now apply the ADHH theorem to the perturbed Ising/Kitaev Hamiltonian $-J\hat{N} + \hat{Y}$, with 
open boundary conditions and the operators defined by (\ref{eqn:MZM-L},\ref{eqn:MZM-Y}).
The theorem is applicable for $J/J_0$ sufficiently large, where the energy scale $J_0$ is given in this case by 
\begin{align}
\nonumber
  J_0=\frac{1}{\kappa_0^2}\biggl[e^{\kappa_0} h&+
e^{2\kappa_0} (h_2 +J_{AB}+J_{BA}+J_{AA}+J_{BB}) \\
  &+e^{3\kappa_0}(\text{3-site terms}) + \ldots\biggr]
\label{J0def}          
\end{align}
The $J_{AB}$ term is included in the presence of disorder, where it is the deviation
of $J$ from its mean value. In the disordered case, $J_0$ is defined as the maximum possible value
of the right-hand side of Eq. (\ref{J0def}) for any site in the system, stipulating that the
interactions on the right-hand side of Eq. (\ref{J0def}) touch that site.
The number $\kappa_0$ is chosen so that this
sum is finite. By choosing $e^{\kappa_0}$ as large as possible while satisfying this requirement, we can
maximize the range of $J$ over which the theorem applies. Then the theorem guarantees that there exists a local unitary transformation ${\cal U}$
such that
\begin{equation}
{\cal U} (-J\hat{N} + \hat{Y})\, {\cal U}^\dagger = -J \hat{N} \,+ \, \hat{D} \, +\,
O\Big(e^{-cn_*}\Big) \, ,
\label{thm}
\end{equation}
where $[\hat{N},\hat{D}]=0$ and $n_*$ is given by (\ref{nstardef}). Another way to say this is that the original Hamiltonian $\hat{H} = -J \hat{N} + \hat{Y}$ has an approximately conserved quantity $\mathcal{U}^{\dagger} \hat{N}\mathcal{U}$, which is conserved until times $t_* = O(e^{cn_*})$.

To illustrate the significance of this approximately conserved quantity, let us consider the limit $J_0 \to 0$, which implies $\hat{Y} = 0$ so the Hamiltonian is simply $\hat{H} = -J \hat{N}$. In that case the local unitary rotation $\mathcal{U} = \mathbb{1}$, so the conserved quantity is simply $\mathcal{U}^{\dagger} \hat{N} \mathcal{U} = \hat{N}$. In this limit, excitations (henceforth, ``quasiparticles'') correspond to making one of the terms in \eqnref{eqn:MZM-L} have eigenvalue $-1$, and have no dynamics. The conserved quantity $\hat{N}$ simply counts the number of quasiparticles, which is the eigenvalue of $(L-1-\hat{N})/2$. In fact, for our purposes, it will be sufficient to use the fact that the number of quasiparticles is conserved modulo 2 in this limit; in other words, here there is a conserved $\mathbb{Z}_2$ charge
\begin{equation}
\label{Ftildedefn}
\mathcal{\widetilde{F}} \equiv e^{i\pi\hat{N}/2} = i\gamma_1^B \left(\prod_{j=2}^{L-1} i \gamma_j^A \gamma_j^B\right) \gamma_L^A.
\end{equation}
We emphasize that this is \emph{not} the same as the fermion parity $\mathcal{F}$, which is also a symmetry of any local fermionic Hamiltonian; the latter is instead given by 
\begin{equation}
{\cal F}= \prod_{j=1}^L (i\gamma_j^A\gamma^B_{j}), 
\label{Fdef}
\end{equation}
which differs from $\widetilde{\mathcal{F}}$ in that the boundary Majorana operators $\gamma_1^A$ and $\gamma_L^B$ are also included. The charge $\widetilde{\mathcal{F}}$ thus can be interpreted as the ``fermion parity in the bulk''.

The important result is that as perturbations are added, moving $J_0$ away from zero, there is a continuous deformation of $\hat{N}$, namely $\mathcal{U}^{\dagger} \hat{N} \mathcal{U}$, which we continue to identify as the effective quasiparticle number, and is approximately conserved. By the intuitive discussion in Section \ref{sec:conceptual}, this suggests that topological information stored in the Majorana zero modes should have a long decoherence time. We are now in a position to rigorously establish this. Indeed, a sufficient condition to allow for quantum information to be stored with infinite (respectively, very large) decoherence time is that there exist Majorana operators $\Psi^A$ and $\Psi^B$ which square to the identity, anti-commute with each other, and commute (respectively, almost commute) with the Hamiltonian; this means that they form an ``edge strong MZM'' in the language of Ref.\ \onlinecite{Alicea15}.
 We will show that this is an implication of the ADHH result.
 
The presence of the strong MZMs has important ramifications.
In the fermionic picture, both $\Psi^A$ and $\Psi^B$ toggle between {\em all} states in the sectors, even the highly excited ones. Their presence means that the full Hilbert space of the system can be decomposed into the tensor product
of a two-state quantum system, i.e.\ a topological qubit, and a non-topological ``bulk'' Hilbert space,
of dimension $2^{L-1}$, such that the Hamiltonian vanishes upon projection onto the topologial qubit,
up to finite-size corrections $\sim e^{-L/\xi}$.
Consequently, the topological qubit is protected at any temperature: regardless of how the dynamics of the
system affects the projection of the state of the system into the bulk Hilbert space, the topological qubit is unaffected.
In the spin language, this means that the autocorrelator of the boundary spin operator $\sigma^z_1=\gamma^A_1$ for any temperature or initial state is non-vanishing up to exponentially long times of order $(J/h)^L$.\cite{Kemp17} The same goes for the other edge spin $\sigma^z_L={\cal F}\gamma^{B}_L$.

To see why ADHH implies a strong MZM, let us first return to the limit ${J_0}=0$ where the Hamiltonian is simply $-J\hat{N}$. In this limit the Majorana operators $\gamma_1^A$ and $\gamma_L^B$ already form a strong MZM. Now turn on any or all of the other terms in $\hat{Y}$, while keeping $0<{J_0}\ll J$.
The ADHH theorem states that there exists a
local unitary change of basis ${\cal U}$ which transforms the problem into one in which $\hat{N}$ is
conserved, up to nearly exponentially small corrections.
The locality properties of ${\cal U}$, along with the fact that $[\hat{N},\hat{D}]=0$,
require that the approximate transformed Hamiltonian $-J\hat{N} + \hat{D}$ commutes with the
edge Majorana fermions:
\begin{equation}
\left[(-J\hat{N} +  \hat{D}), \gamma^A_{1}\right] = \left[(-J\hat{N} +  \hat{D}), \gamma^B_{L}\right] = 0
\label{Dedgecomm}
\end{equation}
To see this, observe that $-J\hat{N} + \hat{D}$ commutes with both the fermion parity $\mathcal{F}$ and the ``bulk fermion parity'' $\widetilde{\mathcal{F}}$ defined in \eqnref{Ftildedefn}. Therefore, it also commutes with their product $\mathcal{F} \widetilde{\mathcal{F}} = i \gamma^A_1 \gamma^L_1$. The fact that $\mathcal{U}$ is a local unitary ensures that the norm of any term coupling both $\gamma_1^A$ and $\gamma_L^B$ must be exponentially small in $L$. We assume that $L$ is large enough that such terms can be ignored. It then follows that all terms must commute with $\gamma^A_1$ and $\gamma^B_1$ individually, proving (\ref{Dedgecomm}).

Thus, in the presence of interactions we define
\begin{align}
\Psi^{\rm l} = {\cal U}^\dagger \gamma^A_1 {\cal U}\ ,\qquad
\Psi^{\rm r} = {\cal U}^\dagger \gamma^B_L {\cal U}\ .
\end{align}
The vanishing commutator (\ref{Dedgecomm}) and the theorem (\ref{thm}) show that these commute with $H$ up to order $cn_*<L$:
\begin{align}
\label{eqn:pretherm-SZM}
[\hat{H},\Psi^{\rm r}]=O(e^{-cn_*})\ ,\qquad [\hat{H},{\Psi^{\rm r}}]=O(e^{-cn_*}) \ .
\end{align}

By construction $\Psi^{\rm l}$ and $\Psi^{\rm r}$ square to the identity operator and anticommute with ${\cal F}$.
We call such operators, satisfying Eq. (\ref{eqn:pretherm-SZM}),
``prethermal strong Majorana zero modes''. Ref.\ \onlinecite{Kemp17} found evidence for ``almost'' strong zero modes in the special case where $\hat{Y}$ contains only
non-zero $h, {h_2}, {J_2}$.
Here we have established that they are an example of a prethermal strong MZM, confirming the claims made there.

\subsection{Temperature dependence of the lifetime}

Now suppose that the system is at temperature $T$. The dynamics of the prethermal strong
MZM $\Psi^{\rm l}$ will be visible in the retarded Green function
\begin{align}
G(\omega) &\equiv \int_0^\infty dt\, e^{i\omega t}\,
\text{Tr}\!\left\{e^{\beta H}\, {{\gamma}^A_1}(t) {{\gamma}^A_1}(0)\right\}\cr
&= \int_0^\infty dt\, e^{i\omega t}\,
\text{Tr}\!\left\{e^{\beta H}\, {\sigma^z_1}(t) {\sigma^z_1}(0)\right\}\ .
\end{align}
This is not the Green function of $\Psi^{\rm l}$ but rather that of the ``bare''
operator ${\gamma}^A_1$. However, there will be non-zero overlap between these two operators,
so the Green function of the latter can be used to probe the former.
In the limit of large system size we can ignore the interaction
between $\Psi^{\rm l}$ and $\Psi^{\rm r}$, so this Green function takes the form
\begin{equation}
G(\omega) = Z(\omega + i\Gamma)^{-1}\ .
\end{equation}
The ``wavefunction renormalization'' $Z$ is a measure of the overlap between $\gamma_1^A$ and
the MZM operator $\Psi^{\rm l}$, as determined by the unitary transformation $\mathcal{U}$.
The decay rate $\Gamma$ will be our primary figure of merit in judging prethermal strong zero modes.
It is directly reflected in the width of a zero-bias peak observed in tunneling into the end of a Majorana chain
(see Sec. \ref{sec:quantum-dot-chains}) and determines the error rate for topological qubits encoded in the MZMs ${\Psi^\text{l}},{\Psi^\text{r}}$. The decay rate $\Gamma$ is determined by $\hat{E}$,
the correction term in the transformed basis, and would vanish if $\hat{E}$ were to vanish, with the decay time becoming infinite. 
From Eq. \ref{eqn:pretherm-SZM}, which expresses the fact that $\| \hat{E} \| = O(e^{-c n_*})$,
we see that
\begin{equation}
\Gamma = A(T)\,e^{-c n_*} \approx A(T)\, e^{-\frac{cJ/{J_0}}{{\ln^3}(J/{J_0})}}\ .
\end{equation}

Let us now examine
the temperature dependence in more detail. After a local unitary
rotation by $\mathcal{U}$, the Hamiltonian can be shifted to
[Eq.(\ref{eqn:rotated-H=N+D+E})] as $\hat{H} = -J\hat{M} + \hat{D} + \hat{E}$,
where $\hat{M} = -\hat{N} + c$, with the constant $c$
chosen so that $\hat{M}$ has smallest eigenvalue $0$. 
The term $\hat{D}$ commutes with $\hat{M}$ and can be jointly diagonalized with it,
 so the decay is attributed to
resonant transitions between different $\hat{M}$ sectors induced by $\hat{E}$. These transitions can only happen
between states with approximately the same energy with respect to $J\hat{M} +
\hat{D}$.
The corresponding energy bandwidth 
of the sector
with $\hat{M}=m$ is bounded by
$[E_0 + mJ - mCJ_0,E_0 + mJ+mCJ_0]$ where $E_0$ is
the ground-state energy, and $C$ is some dimensionless constant.
Roughly, this is because a state in this sector differs from the ground state
only in at most $m$ spots, and only those spots can contribute to the energy difference from the ground state. For a rigorous proof, see Appendix \ref{appendix:energy_bandwidth}.
States in the $\hat{M}=m$ sector can have the same energy as those in the $\hat{M}=m+1$ sector
when $E_0 + mJ+mCJ_0 \geq E_0 + (m+1)J - (m+1)CJ_0$ or, in other words, when
$m \geq m_{\mathrm{min}} \sim J/J_0$. A transition can only occur when the resonance condition is satisfied.
In an excited state with a non-zero density of excitations, at least $m_{\mathrm{min}}$ of them must be close
to the MZM in order for a transition to occur. Hence, 
\begin{equation}
\label{eqn:decay-rate-general}
\Gamma_\text{el-el} = {A_0}\, 
e^{-c n_*} \rho^{m_{\mathrm{min}}},
\end{equation}
where $\rho$ is the density of excited quasiparticles. 

We thus have shown that the decay time of a MZM is of Eq. (\ref{eqn:decay-rate-general}) is determined by two factors.  The important consequence is that decay of an MZM is suppressed by a high power of the
quasiparticle density, in addition to the state-independent
exponential suppression $e^{-c n_*}$.

At very low temperatures,
the density of excited quasiparticles in superconductors is generally higher
than the expected thermal (or pre-thermal) equilibrium value $\rho \sim e^{-\Delta/T}$.
(The reasons for this lie outside the purview of our discussion; for a recent theoretical analysis, see Ref. \onlinecite{Bespalov16}.  References 1-7 in this paper
contain experimental measurements of the quasiparticle density.)
At temperatures that are not too low, the quasiparticle density
exhibits equilibrium behavior, and we have $\rho \sim e^{-\Delta/T}$.
When this is the case, the decay rate is controlled by $e^{-\Delta_{\mathrm{eff}}/T}$,
where we have defined the ``effective gap'' $\Delta_{\mathrm{eff}} \sim
J(J/J_0)$. The effective gap is much larger than the actual gap, which is 
$\Delta \sim J$, up to corrections of
order $J_0$. Therefore, for $J \gg J_0$ and $T \ll \Delta_{\mathrm{eff}}$
the decay rate is of the form
\begin{equation}
\label{eqn:decay-rate-T}
\Gamma_\text{el-el} = {A_0}\, 
e^{-c n_*} e^{-\Delta_{\mathrm{eff}}/T},
\end{equation}
This shows that the finite lifetime of an MZM is exponentially large in
$n_* \sim (J/{J_0})/{\ln^3}(J/{J_0})$
and also exponentially large in $1/T$.
Eq.~\ref{eqn:decay-rate-T} suggests that MZM qubits can be optimized by maximizing $J/J_0$
even at the cost of reducing $\Delta$, since in any case $\Delta_{\mathrm{eff}}
\gg \Delta$ when $J/J_0$ is large.

\subsection{Numerical Results for Prethermal MZMs.}

\begin{figure}[ht]
  \includegraphics[width=3.5in]{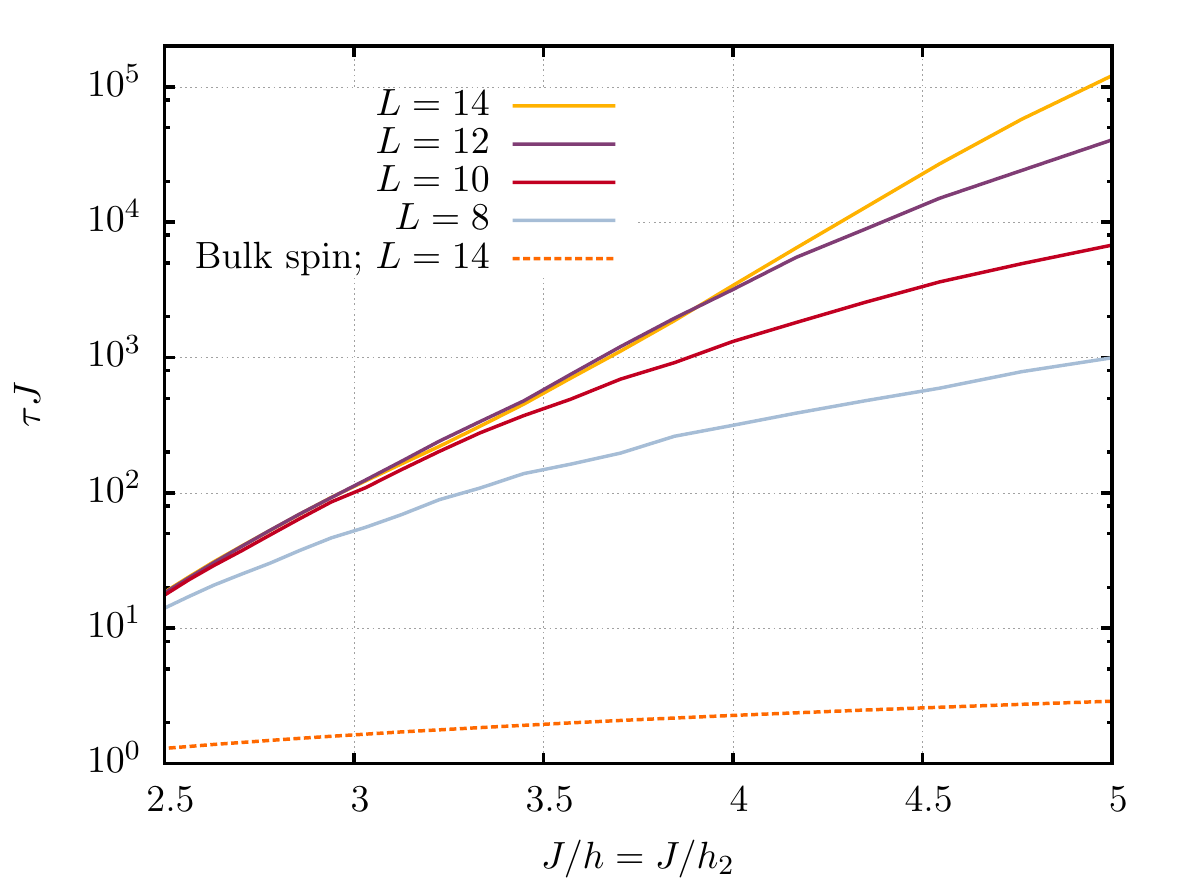}
  \caption{The decay time (on a logarithmic scale) of an MZM at $T=\infty$ for $L=8-14$ sites, as a function
of $h$ and $h_2$, as obtained by exact diagonalization. The data collapses onto the form
$\ln{\tau J} =  J/{h_2}\, f(h/{h_2}) + \text{constant}$, as expected, with deviations due to finite-size effects when the lifetime is long. The decay time of the spin at the centre of the $L=14$ chain (dashed line) is not
prethermalization protected and is much shorter.
  \label{fig:numerics}}
\end{figure}

The preceding general arguments can be substantiated by computations in finite-size systems.
For chains of length $N=8-14$, we study prethermal strong MZMs all the way up to infinite
temperature by exact diagonalization. As may be seen in the top panel of Fig. \ref{fig:numerics},
the MZM survives to very long times at infinite temperature. The lifetimes are consistent with
an exponential dependence on the ratio of scales $J/J_0$ until the lifetime becomes
so long that finite-size effects become important. In a model in which the only terms in $\hat{Y}$
are $h$ and $h_2$, we can write $J_0 = {h_2}/f(h/{h_2})$ for some function $f(x)$.
As may be seen from Fig. \ref{fig:numerics}, the data collapses onto this form.

\begin{figure}[ht]
\includegraphics[width=3.35in]{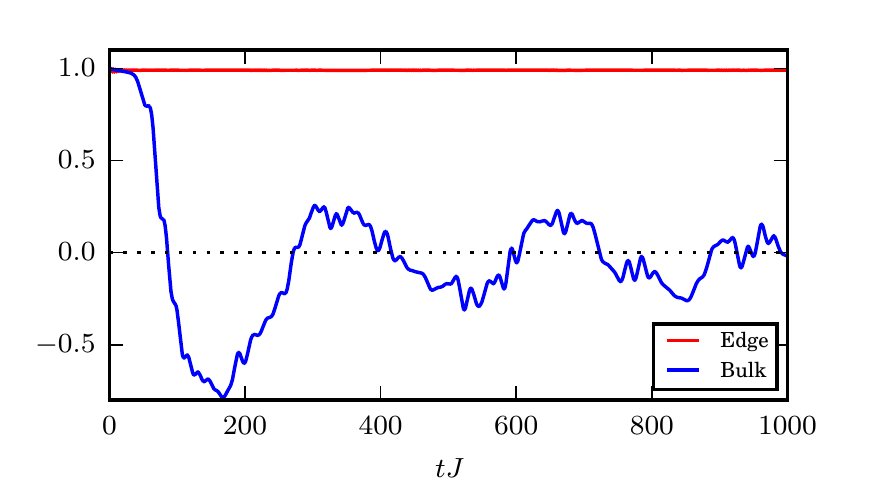}
\caption{\label{fig:spindecay}The decay of an MZM at low energies, as obtained by
TEBD for systems of length $L=40$, which shows that a long-lived MZM persists even in a much larger system.
Red: the decay of the strong zero mode Green's function $\bra{\Psi} 
    \gamma^A_1(t) \gamma^A_1(0) \ket{\Psi}$ for an initial state $\ket{\Psi}$
    containing a single quasiparticle close to the left edge (i.e. $-\gamma^B_3
    \gamma_4^A = -1$ and $\gamma^B_i \gamma^A_{i+1} = 1$ for $i \neq 3$), for
    $L=40$ and $h/J=h_2/J=0.075$. This is equivalent in the corresponding transverse field Ising chain to the spin correlation $\bra{\Psi} \sigma_1^z(t) \sigma_1^z(0) \ket{\Psi}$. Blue: the decay of the bulk spin correlation
    $\bra{\Psi} \sigma_{L/2}^z(t) \sigma_{L/2}^z(t) \ket{\Psi}$ in the transverse
    field Ising chain.}
\end{figure}

Using time-evolving block decimation (TEBD),\cite{Vidal04} we can study much larger systems, where finite-size effects
will be less severe. This approach is applicable to
low-energy initial states for which the entanglement does not grow too much, allowing accurate simulation with bond dimension $\chi=100$.
Our results confirm that, at least for low-energy states, the Majorana
lifetime remains large for larger system sizes (see Figure
\ref{fig:spindecay}). Note that the decay time shown in Figure
\ref{fig:spindecay} should not
be directly compared to Figure \ref{fig:numerics} because Figure
\ref{fig:numerics} is at infinite temperature whereas Figure \ref{fig:spindecay}
is at very low temperature.

\section{General Criteria and Examples for Prethermalization-Protected Topological Degrees of Freedom}
\label{sec:general}

The procedure described in preceding sections uses the ADHH theorem
to find prethermal strong MZMs. In this section, we give a systematic approach
to applying the theorem to determine when topological degrees of freedom
are protected from thermal fluctuations by prethermalization.
This approach involves two key observations:
\begin{itemize}
\item The ADHH theorem guarantees the presence of a single long-lived local U(1) charge for $J/J_0$ large enough.
\item In turn, this nearly conserved U(1) prevents local bulk excitations from violating the conservation of
some topological charge, up to exponentially small terms.
\end{itemize}
More prosaically, the idea is that the nearly conserved quantity guarantees that
a topological charge localized at edges or defects cannot be changed
by the absorption or emission of small numbers of bulk excitations. These general criteria can be applied
in any dimension, and so will give some examples in one, two, and three dimensions.
We illustrate them through several one-dimensional examples in the remainder of this section
and with higher-dimensional examples in the next.

\subsection{Three-state Potts}

It is instructive to analyze an example where the ADHH theorem, although applicable, does {\em not} guarantee any prethermal strong zero modes. As shown in Refs.\ \onlinecite{Fendley12,Jermyn14}, the three-state Potts chain does not have any sort of edge zero mode for finite $J$. The easy edge-spin-flip process that kills the putative zero mode can be seen easily in perturbation theory for $J$ large. Here we rephrase this result in the more general setup of this paper.

The quantum-chain analog of the 3-state Potts model has a three-state quantum system on each of $L$ sites. The basic operators $\sigma$ and $\tau$ acting non-trivially on a single site generalize the Pauli matrices $\sigma^x$ and $\sigma^z$. Instead of squaring to the identity and anticommuting, they obey
\[
\label{z3spin}
\sigma^3 = \tau^3 =1\ , \quad \sigma^\dagger =\sigma^{2}\ , \quad \tau^\dagger = \tau^{2}\ ,\quad
\sigma\tau = \omega\, \tau\sigma\ ,
\]
where $\omega\equiv e^{2\pi i/3}$.  
Matrices satisfying this algebra are
\begin{align}\sigma = 
\begin{pmatrix}
1&0&0\\
0&\omega&0\\
0&0&\omega^2
\end{pmatrix},\quad\quad
 \tau = 
\begin{pmatrix}
0&0& 1\\
1&0& 0\\
0&1&0
\end{pmatrix}\ .
\label{explicitrep}
\end{align}
Here $\sigma$ generalizes the Pauli matrix $\sigma^z$ to measure the value of a clock variable, while $\tau$ generalizes $\sigma^x$ to shifting the value. The operators $\sigma_j$, $\tau_j$ are defined analogously to the Ising case, where they act non-trivially on the $j$th site of the chain and trivially elsewhere.

The three-state Potts chain is invariant under global $S_3$ permutations of the three states and has nearest-neighbor interactions. This fixes the Hamiltonian with open boundary conditions to be $H=-J\hat{N}_{\rm P}+Y_{\rm P}$, where
\begin{align}
\hat{N}_{\rm P} =\sum_{j=1}^{L-1} \left(\sigma_j^\dagger\sigma_{j+1}+\sigma_j\sigma_{j+1}^\dagger\right),\quad
\hat{Y}_{\rm P}= -\sum_{j=1}^L\, \left(\tau_j +\tau^\dagger_j\right) \ .
\label{HPotts}
\end{align}
If desired, these operators can be rewritten in terms of ${\mathbb Z}_3$
parafermionic operators akin to Majorana fermions \cite{Alicea15}.

The operator $\hat{N}_{\rm P}$ has integer eigenvalues, since
$\sigma_j^\dagger\sigma_{j+1}+\sigma_j\sigma_{j+1}^\dagger$ has eigenvalue
$2$ if the ${\mathbb Z}_3$ spins at $j$ and $j+1$ are identical, and $-1$ if they are not.
Thus $\hat{N}_{\rm P}$ is related to counting kinks, just as in Ising. 
However, an important difference with the Ising case is that here there are two types of kinks. We label the three states at each site as $A,B,C$, with the ${\mathbb Z}_3$ symmetry cyclically permuting them. When the states $AB$ (or $BC$ or $CA$, their cyclic permutations) occur on sites $j$ and $j+1$ respectively, we call the configuration a kink, and when $BA$ appears (or $CB$ or $AC$), we call the configuration an antikink. Thus $(2(L-1)-\hat{N}_{\rm P})/3$ counts the number of kinks {\em plus} the number of antikinks.

Since $\hat{N}_P$ is an integer, the ADHH theorem says there is an emergent $U(1)$ symmetry in the prethermal regime, conserving the total number of quasiparticles. However, this symmetry does not prevent a zero mode from decaying, because a kink can scatter off the edge and turn into an antikink without changing $\hat{N}_{\rm P}$.  In perturbation theory this results from the easy-spin-flip process described in Ref.\ \onlinecite{Jermyn14}.
\footnote{A field-theory approach gives a consistent picture. The Hamiltonian (\ref{HPotts}) is not integrable except at the critical point $t=J$, but the corresponding field theory describing the scaling limit is. The resulting exact boundary scattering matrix turns out to always scatter kinks to antikinks\cite{Chim94}.}  Namely, consider say the $AB$ kink on sites $1$ and $2$. The $Y_{\rm P}$ term allows the spin on site $1$ to be shifted from $A$ to $C$, and so converts the $AB$ kink to the $CB$ antikink. This process conserves $\hat{N}_{\rm P}$ while flipping the edge spin.  Clearly there can be no long edge-spin relaxation time, and no edge strong zero mode,
even in the prethermal regime.

This fact appears nicely in the language of the unitary transformations used in this paper. The point is that the transformation guarantees only that the resulting $\hat{D}$ commutes with $\hat{N}$. It says nothing directly about the edge spin, which is why in the Majorana case we needed to argue that no terms involving edge-spin flips could appear in $\hat{D}$. Here they can. Such operators are easy to write out using the projectors
\begin{align}
P^{(r)}_{j,j+1}=(1+\omega^r\sigma_j^\dagger\sigma_{j+1}+\omega^{2r}\sigma_j\sigma_{j+1}^\dagger)/3
\end{align}
satisfying $P^{(r)}_{j,j+1}P^{(s)}_{j,j+1}=\delta_{rs}P^{(r)}_{j,j+1}$ and $\tau_jP^{(r)}_{j,j+1}=P^{(r-1)}_{j,j+1}$. Since $\hat{N}_{\rm P}$ can be written as a sum over $P^{(0)}$, it follows that
\[ \left[ \hat{N}_{\rm P},\, \tau_1P^{(2)}_{1,2}\right] = 0 \ \]
so there is no obstacle to including $\tau_1P^{(2)}_{1,2}$ into $\hat{D}$. Indeed it does appear
for (\ref{HPotts}) and for a generic Hamiltonian with the same dominant term $\hat{N}_P$. Since it shifts the edge spin, it rules out any strong zero mode. The $Q$-state Potts model with $S_Q$ permutation symmetry is the
obvious generalization of this to any integer $Q\ge 2$ and, except for the 
Ising case $Q=2$, the same arguments apply: there is no edge strong zero mode for any finite $J$.

\subsection{${\mathbb Z}_3$ Parafermions}

Despite the results for the three-state Potts model, it is still possible to have prethermal edge zero modes of ${\mathbb Z}_3$ parafermions. One way of doing this is to follow  Refs.\ \onlinecite{Fendley12,Jermyn14} and to deform the dominant term $\hat{N}_{\rm P}$ from (\ref{HPotts}) to
\begin{align}
\hat{N}_{\theta} =\frac{1}{2\cos\theta}\sum_{j=1}^{L-1} \left(e^{-i\theta}\sigma_j^\dagger\sigma_{j+1}+e^{i\theta}\sigma_j\sigma_{j+1}^\dagger\right)\ .
\label{Nchiral}
\end{align}
For $\theta$ not a multiple of $\pi/3$, this explicitly breaks spatial parity and time-reversal symmetries. It also breaks the $S_3$ permutation symmetry to $\mathbb{Z}_3$, and thereby breaks the symmetry between kinks and antikinks.

To utilize the ADHH theorem, we need to chose $\theta$ so that $\hat{N}_{\theta}$ has integer eigenvalues. The simplest non-zero value, $\theta=\pi/3$, is the Potts antiferromagnet.  Here the edge zero mode does not exist for similar reasons as described above for the ferromagnet. Thus we choose $\theta=\pi/6$, halfway between ferromagnet and antiferromagnet. In this case, the energy of an antikink is twice that that of a kink when $h=0$, and $L+\hat{N}_{\pi/6}$ counts the number of kinks plus {\em twice} the number of antikinks. Turning on $h$ and applying the theorem means that the resulting $\hat{D}$ does not conserve the number of kinks individually, but allows scattering processes that convert one antikink to two kinks. The nice fact is that no such simple process flips the edge spin. For example, an antikink near the edge scatters into two kinks via the process
\[BAAAA\dots\quad \Rightarrow\quad  BCAAA\cdots\]
that does not flip the edge spin. More formally, there is no local operator involving $\tau_1$ inside $\hat{D}$ here, just as there is none involving $\sigma^x_1$ in the Majorana case. To prove this, we utilise the identity  \cite{Moran17}
\begin{equation}
\omega^{L + \hat{N}_{\pi/6}} = \sigma_1^{\dagger} \sigma_L\ .
\end{equation}
Hence, conservation of $\hat{N}_{\pi/6}$ implies that any term that does not couple the two ends of the chain must commute with $\sigma_1$ and $\sigma_L$ individually.

The physics for other values of $\theta$ seems akin to the $J_2$ large case discussed next. For certain values of $\theta$, $\hat{N}_\theta$ can be rescaled to have integer eigenvalues, but then edge spin-shifting terms can occur at some order \cite{Moran17}.

\subsection{Interacting Majorana chains with multiple large couplings}
\label{sec:J2large}
We now return to the interacting Majorana chain but allow $\hat{Y}$ to include terms that are not very small. We cannot apply the ADHH theorem to ensure a long decay-time for the zero mode. We may be tempted to fix this by moving the offending terms from $\hat{Y}$ to $\hat{N}$. Naively, this fix would be valid as long as the extra terms commute with our putative zero mode, and the eigenvalues of $\hat{N}$ remain integers. We show that similar considerations as for the Potts and clock models arise: the pre-thermalization theorem holds, but the resulting $U(1)$ symmetry need not protect the zero mode.

As a simple example, returning to our original Ising model in \eqref{eqn:MZM-Y}, let us allow the next-nearest neighbor coupling $J_2$  
to be of the same magnitude as $J$. In order to apply the ADHH theorem, we must then include it in the dominant term $J\hat{N}$. This presents a potential problem, since the theorem requires integer eigenvalues of $\hat{N}$. However, if we take rational $J_2/J = p/q$ for coprime integers $p$ and $q$, then
\begin{align}
\hat{N}_{nn} = \sum_{j} ( q \sigma^z_j \sigma^z_{j+1} + p \sigma^z_j \sigma^z_{j+2}) \ , 
\end{align}
still has integer eigenvalues. The Hamiltonian becomes $\hat{H}_{nn} = -(J/q) \hat{N}_{nn} +\hat{Y}_{nn}$.

We may now use the ADHH theorem to obtain an approximate conservation law for $\hat{N}_{nn}$. However, $\hat{N}_{nn}$ no longer counts kink number; instead it is the sum of the number of broken nearest and next-nearest neighbor bonds, weighted appropriately by $q$ and $p$. This means it is possible to flip the edge spin while conserving $\hat{N}_{nn}$ by converting broken bonds of one type to the other. 

Equivalently, terms will appear in $\hat{D}_{nn}$ which do not commute with the MZM $\gamma^A_1$. For example if $J = J_2$, then $\sigma^x_1 (\sigma^z_2 - \sigma^z_3)$ commutes with $\hat{N}_{nn}$ but flips the edge spin, ruining the conservation of the MZM.
These are the resonances described in detail in Ref.\ \onlinecite{Kemp17}. They allow easy edge-spin flips by exchanging energy between different types of bonds. For example, consider the process which swaps between these two spin configurations, identical but for the edge spin:
\begin{equation}
\uparrow \,\uparrow \,\downarrow\,  \cdots \qquad \Leftrightarrow \qquad  \downarrow \,\uparrow \,\downarrow\,  \cdots \label{eq:domainwallJ}\ . 
\end{equation}
The energy contribution from the three spins on the left is $2(J+J_2)$, while on the right it is $2(2J)$, so when $J=J_2$ the edge-spin can be flipped for no energy cost.

This easy spin flip is analogous to a kink scattering off the edge into an antikink in the Potts model, but
there is an important difference between the two. Here, there is an energy cost associated with either of the domain walls in the right configuration in \eqref{eq:domainwallJ} moving. Thus once a kink has moved to the edge and transformed from the left configuration to the right via an edge-spin flip, it is trapped at the edge, as opposed to the Potts case, where the newly produced antikink is not confined. Thus the only way the kink can move away from the edge is to transform back into the left configuration, reversing the edge-spin flip. This implies that, at low energy densities where there are few kinks, we should expect the MZM to retain a long lifetime, despite the resonances. We have confirmed this via the TEBD.
It should be noted that an exception to this survival at low energy densities occurs at the critical point $J_2 = -J/2$ which, for any non-vanishing $\hat{Y}_{nn}$, becomes a paramagnetic regime between competing ferromagnetic and antiferromagnetic orders. This allows kinks to move freely from the edge in either configuration.

These results and the ADHH theorem show that it is possible for the MZM to survive for long times even for large $J_2/J$, because the term in $D$ with the edge-spin flip may occur only with some high power of $1/(Jq)$. In other words, the order in perturbation theory in which the resonance occurs may be some large value $n_r$. The time to decay will then be order $e^{cn_r}$.  For general $p$ and $q$, we expect that $n_r$ will be roughly max($p,q$), in accord with the analysis of Ref.\ \onlinecite{Kemp17}. It is worth noting that at $J=0$ but non-zero $J_2$, the ensuing Hamiltonian model is equivalent to two copies of the Ising-Kitaev chain, and so $n_*\to\infty$ as $J_2/J\to\infty$ as well.

\section{Two- and Three-Dimensional Systems}
\label{sec:highd}

As we have emphasized throughout, the ADHH theorem applies in any dimension. In this section, we apply the general
criteria developed in Section \ref{sec:general} to show how the resulting almost-conservation law can result in topological protection analogous to the one-dimensional examples we have analyzed.

\subsection{Two dimensions}
\label{sec:2d}

We study a perturbed toric code Hamiltonian \cite{Kitaev97} on a finite square lattice with sides of lengths $L_1$ and $L_2$. The spins live on the
links $i$, with Hamiltonian
\begin{equation}
\label{eqn:toric-code-H}
H = -u\left({\sum_v} A_v + {\sum_p} B_p\right) + h^z \sum_i \sigma^z_i + h^x \sum_i \sigma^x_i + \ldots
\end{equation}
where $A_v = \prod_{i\in \cal{N}({\rm v})} \sigma^z_i$ and $B_p = \prod_{i\in p} \sigma^x_i$ for vertices $v$ and plaquettes $p$.
In the bulk, there are 4 spins on the links entering each vertex $v$ and 4 spins on the links in
each plaquette $p$. We put
``rough'' and ``smooth'' boundary conditions\cite{Bravyi98,Kitaev12}
on, respectively, the horizontal and vertical sides.
At a rough edge, there are only three links around each edge plaquette;
at a smooth edge, there are only three links entering each edge vertex.
At the rough boundaries, $B_p$ is modified so that it is the product
of the three $\sigma^x_i$ operators around a rough boundary plaquette;
the vertex terms are unchanged since there are still four links attached to each vertex (we include no term for the ``dangling'' vertices touching only one link).
At the smooth boundaries, $A_v$ is modified so that it is the product of the
three $\sigma^z_i$ operators around each smooth boundary vertex; the nearby $B_p$
are unchanged since each plaquette still contains four links. The $\ldots$ represents
all other possible local terms, which are assumed to be small, including
a term $\propto ({\sum_v} A_v - {\sum_p} B_p)$ which would give electric and magnetic charges
different energies.

Such a system has a doubly degenerate ground state, which can be used as a qubit. One basis for this qubit is given by the
eigenstates of the electric charge (modulo 2) on a rough edge. To make this more precise, consider the
unperturbed toric code Hamiltonian ((\ref{eqn:toric-code-H}) with all couplings other than $u$ set to zero). Each term in the Hamiltonian is a projector plus a constant, and so ground states are annihilated by each individually. 
Now consider a path ${\cal P}$ of length $P$ stretching along the lattice from one of the dangling vertices on one rough edge to a dangling vertex on the other edge labeled by consecutive links $l_1,\,l_2\dots l_P$ and the operator
$M_{\cal P}=\prod_{k=1}^P \sigma^x_{l_k}$.
Then the eigenstates of the unperturbed toric code Hamiltonian can be grouped into eigenstates of $M_{\cal P}$ with eigenvalues $\pm 1$. It is easy to check that this eigenvalue is independent of the
choice of the path ${\cal P}$. This eigenvalue is the magnetic charge on either smooth boundary,
with which ${\cal P}$ is roughly parallel.
Likewise, the smooth edge corresponds to a rough edge on the dual lattice, and so we can define a path $\widehat{\cal P}$ on the dual lattice stretching from one smooth edge to the other. The electric charge operator is then defined as $E_{\widehat{\cal P}}=\prod_{k=1}^{P} \sigma^x_{l_k}$, and analogously to the magnetic charge, any eigenstate of the Hamiltonian can be grouped into eigenstates of $E_{\widehat{\cal P}}$ with eigenvalue $\pm 1$. However, $E_{\widehat{\cal P}}$ and $M_{\cal P}$ anticommute, since ${\cal P}$ and
$\widehat{\cal P}$ always intersect. There are thus two ground states, not four, and the operators acting on this qubit can be identified as $Z \equiv E_{\widehat{\cal P}}$ and
$X \equiv M_{\cal P}$.

Now consider the perturbed model, in which the other couplings are allowed to be non-zero.
If $h^x$ or $h^z$ is large, the system undergoes a zero-temperature phase transition to a trivial phase; see
Refs. \onlinecite{Trebst07,Dusuel11} and references therein. However, if $h^x$, $h^z$, $\ldots$ are not too large,
then the system will remain in the zero-temperature topological phase, and the ground state will still be
doubly degenerate, up to corrections that are exponentially small in min(${L_1},{L_2}$).
We can still associate the $Z$ and $X$ eigenstates with the eigenstates of electric and magnetic charge
(modulo 2) on, respectively, the rough and smooth edges, but the $Z$ and $X$ Wilson lines, $E_{\widehat{\cal P}}$ and
$M_{\cal P}$, will need to be thickened since the ground
state will have fluctuations in which virtual pairs of $e$ particles straddle ${\widehat{\cal P}}$, and similarly with $m$
particles and ${\cal P}$. The resulting thickened
operators will not be washed out by such quantum fluctuations in the ground state because virtual pairs
never get too widely separated before recombining, so long as the system is in the topological phase. (They
have an opportunity to become more and more widely separated from each other as the
system approaches the quantum phase transition.)

However, conventional wisdom holds that this degeneracy is only a feature of the ground states and the excited states are not
degenerate. Thermally excited pairs of quasiparticles can wander far from each other since they are real,
not virtual, excitations. In particular, the naive expectation is that that the qubit -- meaning the boundary
electric and magnetic charges -- will have a decay rate
$\Gamma$ of order $({L_1} + {L_2})  e^{-u/T}$ at non-zero temperature.
Suppose, however, that $u\gg u_0$, where $u_0$ is built analogously to (\ref{J0def}) from $h^x,\,h^z$ and all other terms lumped into the $\ldots$ remainder.
The interesting question is then whether the prethermal conservation law implied by the ADHH theorem results in the qubit living longer than this naive expectation, analogous to the edge modes in 1D.

Because all $A_v$ and $B_p$ operators mutually commute, their sum has integer eigenvalues. 
The ADHH theorem thus implies that $\hat{N}_\text{TC} \equiv {\sum_v} A_v + {\sum_p} B_p$ is conserved,
up to exponentially small corrections. This integer is zero in the ground state, and otherwise is
simply the number of bulk excitations, not just their parity.
To see if the approximate conservation of the number of bulk excitations protects the qubit, we need to define $X$
and $Z$ operators. The bulk Wilson line operators described above will not work because, even if the number of bulk
excitations is conserved, a bulk excitation can still cross a Wilson line and so flip its charge. However, we can define boundary operators that measure the rough-boundary electric charge and the smooth-boundary magnetic charge by simply pushing ${\widehat{\cal P}}$ to either rough boundary and ${\cal P}$
to either smooth boundary:
\begin{equation}
Z = {\cal U}^\dagger \Big(\prod_{i \in HB} \sigma^z_i \Big) \,{\cal U}\, \, , \hskip 0.5 cm
X = {\cal U}^\dagger \Big(\prod_{i \in VB} \sigma^x_i \Big) \, {\cal U}
\label{Wilson}
\end{equation}
where ${\cal U}$ is the unitary given by ADHH; $HB$ is the set of vertical links that belong to the
horizontal rough boundary at the top of the rectangle (this choice is arbitrary; the bottom would work equally well);
and $VB$ is the set of vertical links that belong to the vertical smooth boundary at the left side of the rectangle.

\begin{figure}
\includegraphics[width=3.5in]{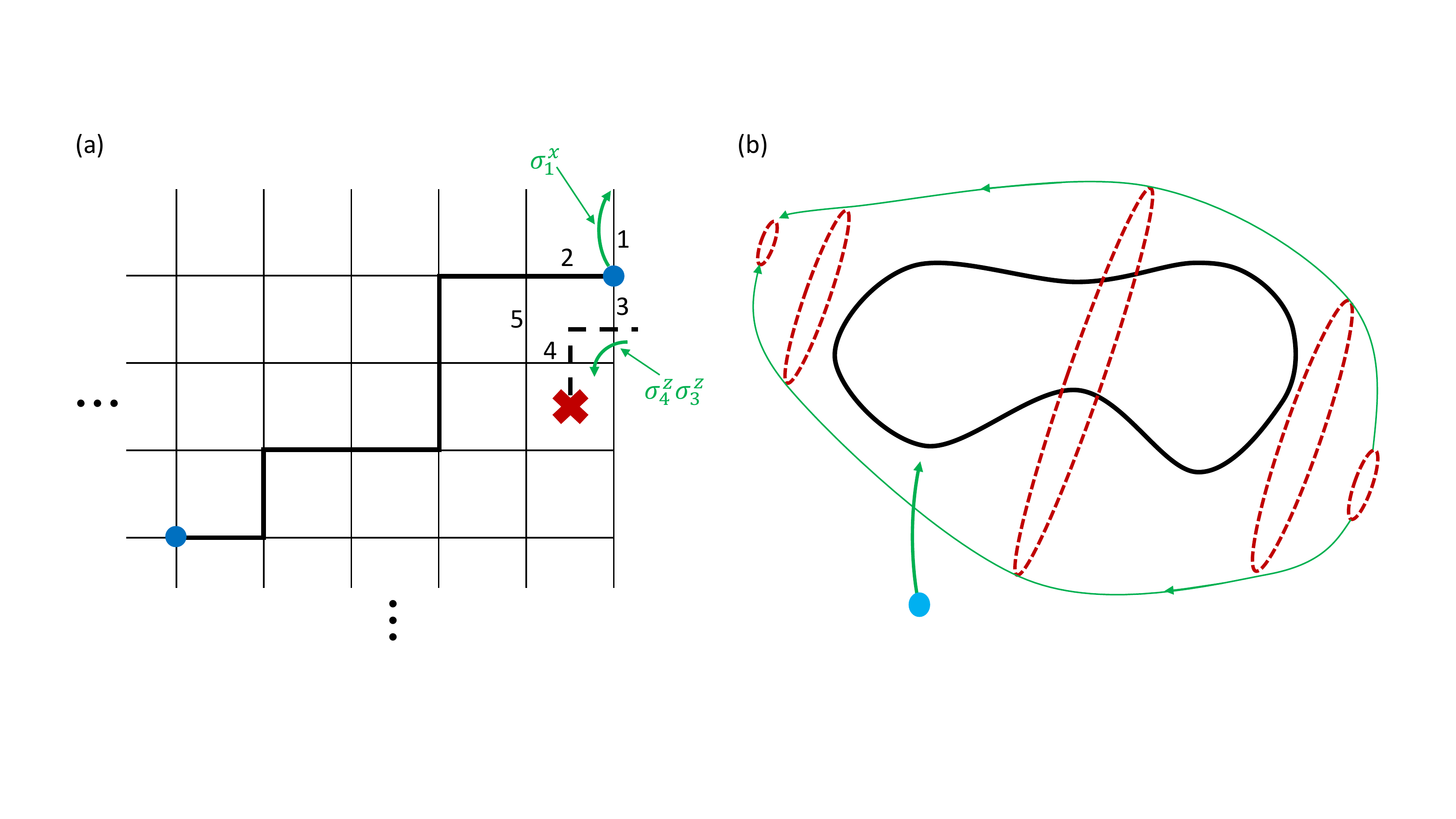}
\caption{(a) Prethermalization does not protect against qubit errors at the boundaries between ``rough'' and ``smooth'' edges
of the toric code on a surface with boundary. Here, an $e$-particle (blue dot) is absorbed by the rough boundary (top edge)
while an $m$-particle (red cross) is emitted by the neighboring smooth boundary (right edge), as described in the text.
(b) When a qubit is encoded in a loop that carries ``Cheshire charge'' in 3D,
errors associated with a magnetic flux loop encircling (sequence of dotted red lines, whose growth and subsequent shrinkage
are indicated by the green arrows) the Cheshire loop are suppressed by the size of the Cheshire loop
since the magnetic flux loops have a line tension at temperatures below the bulk phase transition temperature. 
There is no such protection against the emission or absorption of a point-like electric charge (blue circles),
even below the phase transition temperature into the topological phase. However, prethermalization can suppress such processes exponentially.}
\label{fig:2D-3D-errors}
\end{figure}

The transformed edge Wilson-line operators in (\ref{Wilson}) commute with the Hamiltonian, up to $O(e^{-c n_*})$ corrections,
except at the corners. Away from the corners, $D$ contains only terms commuting with
$Z$, because a non-commuting term would necessarily create or annihilate an $e$ particle in the bulk,
thereby violating the conservation of $\hat{N}_\text{TC}$.
Similarly, terms in $D$ that do not commute with $X$ would create an $m$ particle.  However, at the corners, $D$ can contain terms that cause the rough edge to absorb an $e$ particle and the smooth edge to emit an $m$ particle
or vice versa. One example is $\sigma^z_4 \sigma^z_3 \sigma^x_1 (1+B_{p_{2345}}) (1+A_{v_{123}})$,
where the dangling link is labeled by $1$, 
$v_{123}$ is the corner vertex connected to links $1,2,3$ and $p_{2345}$ is the corner plaquette that overlaps it
and contains links $2,3,4,5$. An error-causing process associated with this term is depicted in Fig. \ref{fig:2D-3D-errors}

We thus arrive at the interesting result that the only possible violations of the long-lived conservation law occur at the corners. Prethermalization therefore suppresses the low-temperature error rate from $\Gamma \propto ({L_1}+{L_2}) e^{-u/T}$ to
\[\Gamma \propto e^{-c {n_*}}({L_1}+{L_2}) e^{-(u/{u_0})u/T}
+ e^{-u/T}\]
for some constant $c$.
It would be interesting to see how this argument generalizes to
weakly perturbed Levin-Wen models \cite{Levin05a}.

\subsection{Three dimensions}
There are four-dimensional topological phases with no point-like excitations,
and they protect quantum information at non-zero temperatures below the phase transition into the topological phase,
as noted in Sec. \ref{sec:conceptual}. In three dimensions, we are halfway there, since magnetic excitations
are loop-like; their line tension prevents them from causing errors. However, there are still point-like excitations
that will cause errors at non-zero temperature. Prethermalization can suppress them.

It was recently shown \cite{Else17b} that Abelian topological phases in $3+1$ dimensions have loop excitations
carrying Cheshire charge. The simplest example is the 3D toric code, which has a Hamiltonian of the same
form as Eq. (\ref{eqn:toric-code-H}) generalized to the cubic lattice, with one modification discussed below.
There are $6$ links attached to each vertex, so $A_v$ is a product of $6$ $\sigma^z_i$
operators while $B_p$ is still the product of
$4$ $\sigma^x_i$ operators around each plaquette. There are loop excitations on which
$e$ particles condense in a manner analogous to the ``rough'' boundaries considered above in the case of
the 2D toric code. To analyze such loop excitations, we can modify the Hamiltonian so that the ground
state has one at a specified loop $K$, as was done in Ref. \onlinecite{Else17b}. We label the loop $K$,
of length $L$, in terms of vertices $v_k$ and links $l_k$ such that the vertices $v_{k+1}$ and $v_k$ are connected by the link $l_k$ for all $k=1\dots L$ with $v_{K+1} = v_1$.
We now modify the Hamiltonian along this loop according to:
\begin{equation}
\label{eqn:condense-e-particles}
\sum_{k=1}^K A_{v_k}\ \rightarrow\ -\sum_{k=1}^K \sigma^x_{l_k}\ .
\end{equation}
This transverse field commutes with the $B_p$ terms, so eigenstates of the Hamiltonian are eigenstates of
$\sigma^x_{l_k}$ for every link $l_{k+1}$ on the loop. Thus they resonate between a state with two $e$ particles at either end of the link and a state without $e$ particles at the ends of the link. Alternatively,
we can insert such a loop excitation in a manner that emphasizes the similarity with the ``rough'' edge
in 2D: we remove all ${v_k}, {l_k} \in K$ from the lattice. The Hamiltonian is unchanged at all of the remaining
vertices and the plaquette operator is modified to a product of 3 spins for the plaquettes that previously
contained ${l_k} \in K$. In this alternative construction, it is again clear that the states of the system
resonate between having zero and two $e$ particles at either end of each link on $K$.
The spectrum is degenerate in the limit of a large loop,
provided that there are at least two such
loops in the system, since the total charge on a loop can be either $1$ or $e$, subject to the constraint that the total
topological charge of the system is fixed. In the simplest case, in which
there are two such loops in the system, states are doubly degenerate and the degenerate subspace at each energy
forms a qubit spanned by states with charge $1$ or $e$ on both loops.
This charge is locally unobservable, hence it is ``Cheshire charge'', which explains why
the two states are degenerate in the limit
of a large loop.

A straightforward generalization of the arguments applied in two dimensions in the section \ref{sec:2d} show that Cheshire-charge-carrying loops cannot emit or absorb an $e$ particle in the analogous prethermal regime in 3D. Hence, the $Z$
operator acting on the qubit commutes with the prethermal Hamiltonian $\hat{D}$. 	A dressed version of the $X$ operator, meanwhile,
is expected to be conserved in the low-temperature phase $T<T_c$ below the phase transition at which long flux loops
unbind and proliferate \cite{Alicki10}. Thus, the qubit is partially protected by the dynamics of the low-temperature phase
(as was already known \cite{Dennis02}) and partially protected by prethermalization. Unlike in the 2D case discussed above, our 3D topological qubit has error rate $\Gamma < e^{-c{n_*}} e^{-(u/{u_0})u/T}$, where $u_0$ is the appropriate energy scale
derived from the couplings in $\hat{Y}$. The corner error processes that caused trouble in 2D are not present here
because an electric charge cannot become a magnetic charge since one is point-like and the other loop-like, unlike
in 2D, where both are point-like and necessarily had the same energy in the prethermal limit.
In other words, in the prethermal regime, the 3D toric code
is, up to exponentially small corrections, a self-correcting non-zero temperature quantum memory;
the only other known examples are
4D topological phases \cite{Dennis02}. Thus, in this case, prethermalization buys us an extra dimension,
which might be rather difficult to otherwise acquire.

\section{Experimental Realizations}
\label{sec:experiment}

Although the primary focus of this paper is the conceptual advance in the theory of topological
phases of matter that results when the theory of prethermalization is brought to bear on it, it is important
to note that this advance may have implications for near-term devices and experiments. For illustrative
purposes, we discuss two examples: Majorana zero modes in semiconductor-superconductor devices and Ising
spins in a trapped ion chain. The former example allows us to introduce an important point, which is that
prethermalization is generically terminated by a coupling to a heat bath; when the coupling is weak, this occurs
at very late times, and the prethermal regime extends until then. Since the electron-phonon interaction
is suppressed by the ratio of the electron mass to the ion mass (it must vanish in the limit that the latter approaches
infinity), this can be weak, thus giving an example of this scenario. Meanwhile, the ion chain has interactions that are long-ranged,
so it is not immediately obvious that our analysis applies. However, the calculations that we present do, in fact, indicate that
prethermalization occurs in this case, as well. Although this system does not have a topological phase, the
edge spins are governed by similar dynamics. We present a quantitative discussion of the experimental requirements
for observing edge spins protected by prethermalization.

\subsection{Quantum Dot Chains and the Fate of Prethermalization in
the Presence of Electron-Phonon Interactions.}
\label{sec:quantum-dot-chains}

Refs. \onlinecite{Sau12,Fulga13} have proposed devices in which quantum dots (possibly defined by gates
acting on MZM-supporting nanowires) are the basic building blocks for super-lattices that have an effective low-energy description as models of Majorana fermions hopping on a lattice. They therefore realize literal versions of the Hamiltonian of Eqs.\ \ref{eqn:MZM-L} and \ref{eqn:MZM-Y}. Moreover,
Ref. \onlinecite{Fulga13} gives a procedure for tuning a system of quantum dots to the limit
in which the coupling $J$ is much larger than all the other coupling constants from
Eq.\ \ref{eqn:MZM-Y}. In the standard notation for the Kitaev chain\cite{Kitaev01} this translates into making $t+\Delta$ much larger than $\mu$ and $t-\Delta$.
In other words, Ref. \onlinecite{Fulga13} gives a procedure for tuning into the
prethermal regime of large $J/J_0$. These devices are a bit futuristic, so we are not, in this section, proposing
immediate experimental tests of prethermalization-protected MZMs. Rather, our purpose here is
to illustrate the considerations that must enter into any effort to exploit prethermalization
in a solid.

As we will discuss momentarily, prethermalization leads us to predict extremely small decay rates for
the MZMs at the ends of such a chain.
However, these extremely small decay rates will not be observed
since the electron-phonon coupling will
end prethermalization before these exponentially small effects do.
In other words, the electron-phonon coupling can
violate the conservation of $N$ before the time $t_*$.
But if the electron-phonon contribution to thermalization is smaller than the electron-electron
contribution over some range of temperatures, device geometries, and device parameters,
then prethermalization will still play an important role in extending the lifetime of a topological qubit.
We will show that the naive MZM decay rate due to electron-electron interactions, which is what would
be seen outside the prethermal regime, is $\Gamma_\text{el-el}^\text{non-pt} \sim n_\text{qp}^2 \cdot 0.6\, \text{GHz}$.
Meanwhile, the decay rate due to electron-phonon interactions is
$\Gamma_\text{el-ph} < n_\text{qp} \cdot 10\, \text{MHz}$, which is almost certainly a gross overestimate
since it doesn't take into account the screening of piezo-electric interactions by the superconductor or the
effect of the device geometry,
which can be designed to have a phonon band gap. Since the former is quadratic in $n_\text{qp}$
while the latter is linear, we conclude that, even for this overestimate of phonon effects,
the electron-electron interactions are the dominant decay channel for $n_\text{qp}> 0.01$, and
prethermalization suppresses this dominant
decay channel. If the bulk quasiparticle density
is thermal (which may or may not be a valid assumption, for the reasons
discussed in Ref. \onlinecite{Bespalov16}; and references 1--7 of that paper),
this translates into a range of temperatures $(t+\Delta)/T < 4.6$. Again, this range could be greatly increased, depending on the effects of
device geometry on the phonon spectrum.
In the rest of this subsection, we explain these estimates in more detail.

The initial motivation for considering such a system was that
it may be possible to tune parts of the system
more reliably into and out of the topological superconducting phase.
The potential drawback of such systems is that the energy gap of the coupled-MZM system
is significantly smaller than the energy gap of a single nanowire. For instance,
Ref. \onlinecite{Fulga13} finds $t \approx 9 \mu$V and $\Delta\approx 6\mu$V.
One might consequently fear that much lower temperatures would be necessary in order
to protect the end MZMs of such a quantum-dot chain.

Indeed, suppose that the hopping parameter were twice as large, in which case the system would be outside the
prethermal regime, with $(t+\Delta)/2 = 7.5 \mu$V and $(t-\Delta)/2 = 1.5 \mu$V,
and assume a nearest-neighbor repulsion $V=3 \mu$V.
Then, a rough naive estimate for the decay rate due to electron-electron interactions would be
$\Gamma_\text{el-el}^\text{non-pt} \sim n_\text{qp}^2 \cdot {V^2}/(t-\Delta)$, since
$(t-\Delta)$ is the bandwidth of excited quasiparticles (so its inverse is the
density-of-states). This gives $\Gamma_\text{el-el}^\text{non-pt} \sim n_\text{qp}^2 \cdot 0.6$GHz.
Thus, the zero mode will decay in a few nanoseconds unless the quasiparticle density is very low,
which is unlikely to be the case since the gap is relatively small.

However, the results of this paper show that the small gap may not doom the MZM:
prethermalization will protect quantum information until a time that is nearly exponentially long in
the ratio between $t+\Delta$ and some combination of the other couplings, such as $t-\Delta$, $V$, etc.,
according to Eq. \ref{eqn:decay-rate-T}. Moreover, the temperature
dependence has a characteristic energy scale $n_* (t+\Delta)$,
rather than $(t+\Delta)$ itself, so relatively small $(t+\Delta)$ is not as detrimental as one might fear.
As a result of prethermalization, the decay rate due to electron-electron interactions should,
instead, be $\Gamma_\text{el-el}^\text{pt} \sim (n_\text{qp})^{m_\text{min}} \cdot
e^{-c n_*}\cdot{V^2}/(t-\Delta)$, as in Eq. (\ref{eqn:decay-rate-general}). In other words,
prethermalization suppresses the decay rate due to electron-electron interactions by a factor
$\Gamma_\text{el-el}^\text{pt}/\Gamma_\text{el-el}^\text{non-pt} \sim
(n_\text{qp})^{m_\text{min}-2} \, e^{-c n_*}$.
For the values of $t, \Delta$ given above, $(t+\Delta)/(t-\Delta) \approx 5$ and
assuming that all other couplings are smaller than $t-\Delta$,
we find $n_* \approx m_\text{min} \approx 5$.
If we replace the $O(1)$ constant $c$ by $1$, then we can use 
$e^{-n_*} \approx 6\times10^{-3}$. Thus, we have
$\Gamma_\text{el-el}^\text{pt} \sim n_\text{qp}^{5} \cdot 3.6$MHz.

If we assume that the system is at a temperature $T= 50$mK,
and that the quasiparticle density is given by the equilibrium value,
then $n_\text{qp} \approx e^{-(t+\Delta)/T} \approx 0.05$ while
$(n_\text{qp})^5 \approx e^{-\Delta_\text{eff}/T} = e^{-m_\text{min} (t+\Delta)/T}
\approx 3 \times 10^{-7}$.
(We are assuming that, in the initial state, the quasiparticle density is
equal to its equilibrium value but its subsequent evolution
is impeded by prethermalization, especially its equilibration with the edge modes.)
Thus, the enhancement of $(t+\Delta)$ to
$\Delta_\text{eff}=m_\text{min} (t+\Delta)$ is the larger effect at these temperatures, assuming that the system
equilibrates rapidly with respect to $\hat{D}$ i.e. is in the prethermal state.
The resulting decay rate is $\Gamma_\text{el-el}^\text{pt} \sim n_\text{qp}^{5} \cdot 3.6$MHz $\sim 1$Hz.
However, if the system
is out-of-equilibrium, the factor of $e^{-cn_*}$ may be more important since it will protect the zero modes
even if there are non-equilibrium excitations in the bulk which would render the equilibrium
estimate $(n_\text{qp})^{m_\text{min}}e^{-\Delta_\text{eff}/T}$ moot.

We must compare the prethermal decay rate to the phonon-assisted decay rate 
$\Gamma_\text{el-ph}$.
A bulk fermionic excitation, which has energy $E=t+\Delta$,
can be absorbed by a zero mode and its energy can be emitted as a phonon of momentum $q=E/v$, where
$v$ is the speed of sound.
The Hamiltonian governing the electron-phonon interaction is:
\begin{equation}
H_\text{el-ph} = \int {d^3}x {d^3}x' \rho_\text{el}(x) V_{ij}(x-x') {\partial_i}{u_j}(x')
\end{equation}
Here, $\rho_\text{el}(x) = i\gamma_1^A \gamma_1^B |{\psi_1}(x)|^2$, where
${\psi_1}(x)$ is the wavefunction of the zero-energy fermionic level of a single dot coupled to a superconductor
\cite{Fulga13}; ${u_j}(x)$ is the displacement in the $j$-direction
of the ion whose equilibrium position is $x$; and $V_{ij}(x-x')=D\delta(x-x')\delta_{ij} + eh_{14}w_{ij}(x-x')$.
The electron-phonon coupling has two parts, the deformation potential $D$ and the piezoelectric coupling $h_{14}$.
The piezo-electric potential satisfies ${q_i}w_{ij}(q)={\sum_\lambda}i{M_\lambda}(q){\epsilon^\lambda_q})_j$
where $\lambda$ are the phonon polarizations, $\epsilon^\lambda_q$ are the corresponding polarization unit
vectors; and ${M_\lambda}(q)$ depend on the direction of $q$ but do not its overall scale.
Hence, the decay rate for an MZM due to the deformation potential electron-phonon coupling is
\begin{align}
\Gamma^{\text{DP}}_\text{el-ph} &= \int \frac{{d^3}q}{(2\pi)^3} |Q(q)|^2 (Dq)^2 \,\frac{1}{\rho}\,
\delta({E^2} - {v_l^2}{q^2}) \, n_\text{qp} \cr
&< \frac{1}{4\pi^2 \rho v_l E } \left(D\, \frac{E}{v_l}\right)^2 \left(\frac{E}{v_l}\right)^2
n_\text{qp}
\end{align}
Here, $\rho$ is the density of the solid and, as before,
$n_\text{qp}$ is the probability of a bulk fermionic excitation on
dot $1$. In going to the second line,
we have bounded $Q(q) \equiv \int {d^3}x\, e^{iq\cdot x} |{\psi_1}(x)|^2$
by $|Q(q)|^2 <1$. The first two factors in the second line
are the matrix element for such a process; the third factor in the second line
is the density of states for the phonon, which is $\propto q^2$;
and the final factor is the probability for a bulk quasiparticle excitation to be near
enough to the MZM for absorption
to occur. (We emphasize that the relevant gap for phonon-assisted decay of an MZM
is $E=t+\Delta$, not $\Delta_\text{eff}$, which is the relevant scale for electron-electron
interactions in the pre-thermal regime.)
The reverse process, in which a bulk quasiparticle excitation
is emitted and a phonon is absorbed, has the same amplitude at low-temperature.
Prethermalization
will occur if $\Gamma_\text{el-ph} < \Gamma_\text{el-el}$.
For InAs we take the following values\cite{Vurgaftman01}: $D=5.1$ eV; the speed of longitudinal sound waves is ${v_l} \approx 4.7$ km/s;
the density is $\rho\approx 5.67$ g/cm$^3$.
We take $t+\Delta \approx 15 \mu$V estimated in Ref. \onlinecite{Fulga13}, as in our discussion of
the prethermal decay rate due to electron-electron interactions. This gives
$\Gamma^{\text{DP}}_\text{el-ph} < n_\text{qp} \cdot 300 $kHz.
For an equilibrium distribution of excited quasiparticles at $T = 50$mK,
this gives $\Gamma^{\text{DP}}_\text{el-ph} < 15 $kHz.
Turning now to the piezoelectric coupling, we first note that,
in the presence of strong coupling to superconducting leads, this effect may be suppressed by screening.
However, if we neglect this screening effect and compute, as an upper bound, the decay rate due to an unscreened piezoelectric coupling, we find:
\begin{align}
\nonumber
\Gamma^{\text{PE}}_\text{el-ph} &< \int \frac{{d^3}q}{(2\pi)^3} |Q(q)|^2 (eh_{14})^2 \,\frac{1}{\rho}\,
\delta({E^2} - {v^2}{q^2}) \, n_\text{qp}\\
&\sim \frac{1}{4\pi^2 \rho v E } (eh_{14})^2 \left(\frac{E}{v}\right)^2
n_\text{qp} .
\end{align}
Here, we have made the approximation of ignoring the difference
between the longitudinal and transverse sound velocities
${v_l}\approx 4.7 $km/s and ${v_t}\approx 3.3 $km/s and simply set them both to $v\equiv 4.2 $km/s.
We have also made the simplification of replacing ${M_\lambda}(q)$ by an upper bound ${M_\lambda}(q)<1$.
Using $h_{14} = 3.5\times10^6 $V/cm, given in Ref. \onlinecite{Madelung04},
we find $\Gamma^{\text{PE}}_\text{el-ph} \sim n_\text{qp}\cdot 10 $MHz.
For an equilibrium distribution of excited quasiparticles at $T = 50$mK,
this gives $\Gamma^{\text{PE}}_\text{el-ph} \sim 500 $kHz.
We note that this estimate does not take into account the effect of phonons in the superconductor,
although we expect this to be a smaller contribution to the decay rate since the electronic wavefunction
is concentrated primarily in the semiconductor; it has also not taken into account the
effect of the device geometry on the phonon spectrum at the wavelengths of interest.
In this regard, we note that it may be possible to pattern a material in order to engineer
a phonon band gap at the wavelength $2\pi k^{-1} = hv/E \approx 1\mu$m,
potentially strongly suppressing the effect of phonons. Thus, even at relatively high temperatures,
$1 \mu$s is a conservative estimate of the potential lifetime of a MZM in a quantum-dot chain in
the prethermal regime, but the lifetime may be as long as $1$ ms.

We briefly note that much of our discussion of quantum-dot chains
in Sec. \ref{sec:quantum-dot-chains} also applies to the model of Ref. \onlinecite{Barkeshli15},
which uses short topological nanowires to construct
a two-dimensional model of Ising anyons on the honeycomb lattice \cite{Kitaev06a}.
The Ising anyons emerge as low-energy
excitations of a superlattice of Coulomb-blockaded islands containing, each containing two nanowires.
Our results do not apply to the physics of the nanowires themselves, but instead to
the effective model of the low-energy degrees of freedom, which has a 
pre-thermal regime in the limit that the bulk Majorana fermion operators have a flat band.
The existence of such a prethermal regime would facilitate universal topological quantum computation
using the strategy of Ref. \onlinecite{Barkeshli15},
since it ameliorates the drawback of a reduced energy gap.

\subsection{Trapped Atomic Chains}
Another possible experimental realization would be a trapped ion or neutral atom chain governed by a perturbed transverse field Ising model \cite{Neyenhuis16, Hild14}. Here, coupling to an external heat bath would be less of a concern, although the effective system size might be smaller than in the quantum-dot case.

For example, in Ref.\ \onlinecite{Hess17}, the authors use chains of up to 22 $^{171}\mathrm{Yb}^+$ ions in linear radiofrequency (Paul) traps, encoding effective two-state systems in their $^2\mathrm{S}_{1/2}$ hyperfine ground states. Long-range spin-spin interactions are generated using laser-mediated spin-phonon interactions. In particular, using the beatnote between two overlapped laser beams to drive stimulated Raman transitions, they generate the effective Hamiltonian
\begin{equation}
  \hat{H} = \sum_{i<j} J_{i,j} \sigma^z_i \sigma^z_j +B \sum_i \sigma^x_i \ ,
\end{equation}
where interaction is long-ranged and antiferromagnetic:
\begin{equation}
  J_{i,j} =  \frac{J_I}{|i - j|^\alpha}\ .
\end{equation}
with $J_I>0$. For nearest-neighbor interactions in Ising, ferromagnet and antiferromagnet are unitarily equivalent. Here the distinction is important, because the ferromagnet has a phase with long-ranged order for $\alpha < 2$ and non-zero temperatures less than
some critical temperature $T_c$,\cite{Dutta01} while the antiferromagnet does not have such an ordered phase for $T>0$, like the nearest-neighbor model \cite{Kerimov99}. Consequently, for initial states that are near the \emph{top} of the spectrum here, the end and bulk spin lifetimes will be infinite since these are low-energy state of the ferromagnetic Hamiltonian $-H$.

In the setup of Ref.\ \onlinecite{Hess17}, the experimentally realizable range of $\alpha$ is $0.5 < \alpha < 2$, while $J_I/2 \pi \leq 1$kHz, achieved by changing the trap voltages and the detuning of the beatnote from resonance. The $B$ term is generated by driving further resonant stimulated Raman transitions out of phase with the beatnote. It can range from negligible to a maximum of $ B/2\pi = 10$kHz. So it should certainly be feasible to enter the prethermal regime $B \ll J_I$.

The autocorrelators of individual spins $\langle\sigma^z_j(t)\sigma^z_j(0)\rangle$ may be measured up to times of order $100/J_I$. Thus, it should be possible to observe the prethermal protection of the edge spin through the survival of its autocorrelator.  In contrast, the bulk spins decay over experimentally accessible timescales. The one caveat is that, although the edge spin will be long-lived for any system temperature or any energy initial state, for very high-energy initial states near the top of the spectrum (or for negative temperatures), the protection comes from the long-ranged ferromagnetic order when $\alpha < 2$, rather than prethermalization. This case is easily distinguishable because the bulk spins will also be long-lived in such initial states.

\begin{figure}[ht]
  \includegraphics[width=3.25in]{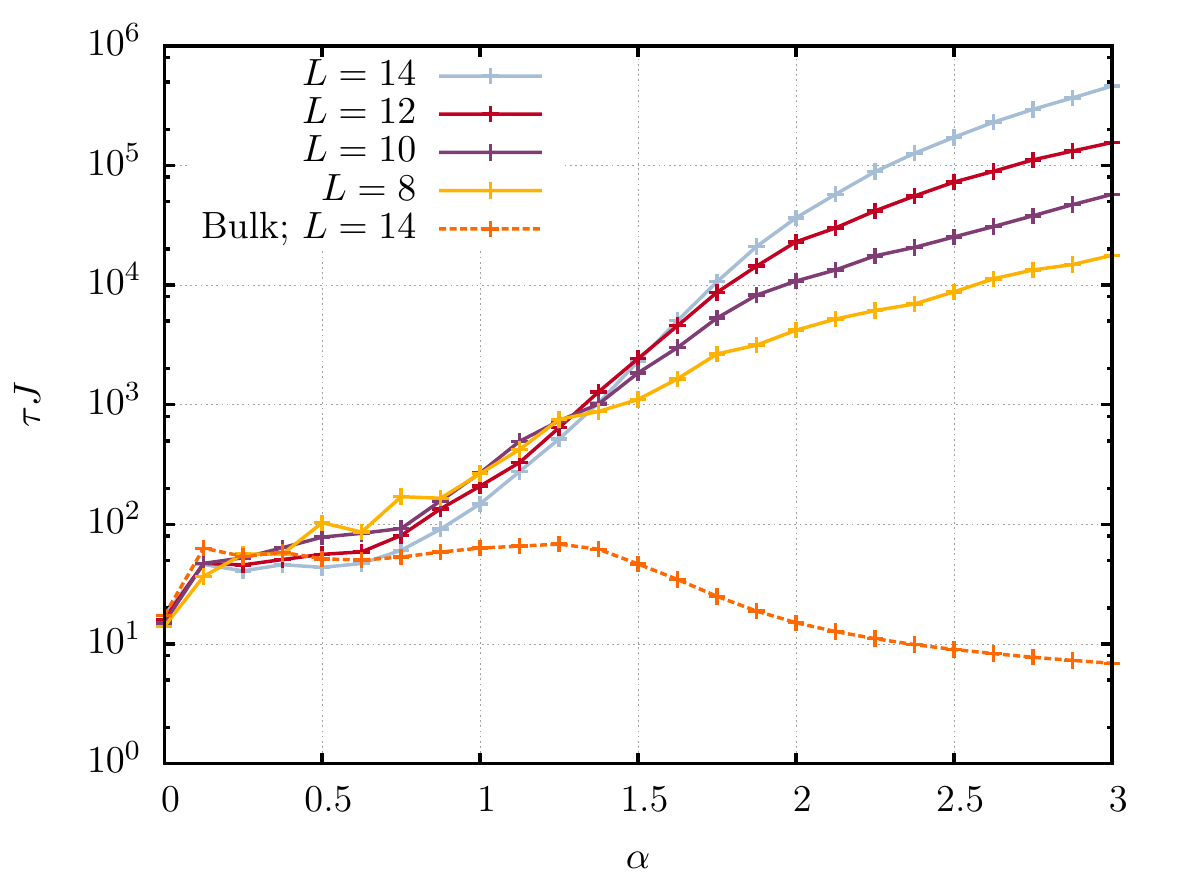}
  \caption{The decay time (on a logarithmic scale) of an MZM in the transverse-field Ising model with long-ranged interactions at $T=\infty$ for $L=8-14$ sites and $h = 0.2$, as obtained by exact diagonalization. It is plotted as a function of $\alpha$, the power of the decay of the long-ranged interaction. As in Fig.~\ref{fig:numerics}, the decay time saturates in $L$ for larger values of the perturbing couplings. The decay times for ferromagnetic and antiferromagnetic interactions are the same at $T=\infty$; it is only at lower temperatures where the effect of the long-ranged order for $\alpha<2$ in the ferromagnetic case becomes important.
  \label{fig:longrange}}
\end{figure}
Simulations using exact diagonalization at infinite temperature on similar system sizes confirm this picture, at least for $\alpha \gtrsim 1.25$, see Fig.~\ref{fig:longrange}.
The main theoretical concern for prethermalization is the long-ranged nature of the interaction, because for the ADHH theorem to hold we require small $J_0/J$. In fact the ADHH theorem has not yet been proven for power-law decay interactions. In our discussion above, $\hat{N}$ consists of just the nearest-neighbor interaction magnitude $J_I$. For $\alpha = 2$, the shortest-possible range in the experiments, the next-largest term is a quarter of the size, so we might be justified in putting it and longer-range terms in $\hat{Y}$. However, for smaller $\alpha$ we should include at least the next-nearest-neighbor term in $\hat{N}$ as well. This case, including the possibility of resonances, is discussed in detail in Section~\ref{sec:J2large} below. Crucially, for there to be any chance for the ADHH theorem to hold, $\hat{N}$ must have integer eigenvalues, which is of course impossible to tune exactly experimentally for more than one coupling. However, for the system sizes $L\leq 22$ which are experimentally accessible, we do not expect this to be of practical issue in observing the prethermal protection of the edge spin in contrast to the bulk, and indeed the exact diagonalization results support this.

Of course, once $\alpha$ becomes small, the arguments above based on locality break down completely. The two ends of the chain come into contact, so that terms which flip the edge spin may be immediately added to $\hat{D}$, and the edge spin will no longer be protected, regardless of the applicability of the ADHH theorem. It is also worth noting that the smaller $\alpha$, the less localised at the edge the zero mode is, and so the greater overlap it has with the bulk spins. This means that there may be some small but observable part of the bulk spin correlator which survives to long times, albeit exponentially suppressed in magnitude compared to the edge.

\section{Integrable Systems}
\label{sec:integrable}

Turning now to more formal considerations, we emphasize that
the number $n_*$ following from the ADHH theorem only defines a lower bound on how long a zero mode will live. 
Indeed, in a free-fermion system, the recursive procedure
defined in Ref. \onlinecite{Abanin15b} converges, so that $n_* \to\infty$ as $L\to\infty$, as we now show explicitly.
In fact, one might go so far as to say that the ADHH recursive procedure gives us an alternate formulation of Onsager's solution
of the Ising model.

The strong MZMs and corresponding unitary operators in the free-fermion Ising/Kitaev chain can be found directly, with no need for the full-blown ADHH procedure. The Hamiltonian here is $H=-J\hat{N}+\hat{Y}$ from 
(\ref{eqn:MZM-L}), and (\ref{eqn:MZM-Y}) with all couplings other than $h$ and $J$ set to zero.
The strong MZMs at the left and right edges respectively are then given by \cite{Kitaev01}
\[
\Psi={\cal N}\sum_{j=1}^L \left(\frac{h}{J}\right)^j\gamma_j^A\ ,\quad
\Psi'={\cal N}\sum_{j=1}^L \left(\frac{h}{J}\right)^j\gamma_{L+1-j}^B\ .
\]
where the normalization ${\cal N}$ is chosen to make $\Psi^2=(\Psi')^2=1$. Each of these commutes with the Hamiltonian, up to terms of order $(h/J)^L$. Thus $n_*$ for the edge modes is $L$ throughout the ordered phase $h<J$. The unitary operator ${\cal U}_e$ relates the strong edge MZM at arbitrary $h$ to that at $h=0$ (the latter commuting with $\hat{N}$. We have
\begin{equation}
  \Psi = {\mathcal U}_e^\dagger \gamma^A_1 {\mathcal U}_e\ ,
\end{equation}
and we write
\begin{equation}
  {\mathcal U}_e = \exp (i [U_0 +U_1 + U_2 + \ldots]),
\end{equation}
where $U_n$ is Hermitian and of order $(h/J)^n$.
If we insist that $\Psi$ commute with the Hamiltonian $\hat{H}$ up to order $(h/J)^n$, as per Section~\ref{sec:prethermal}, then we may calculate $U_n$ by inverting the equation:
\begin{equation}
[\hat{H}, [U_n, \gamma^A_1]] = i [\hat{H},  e^{-i \sum_{j=0}^{n-1} U_j} \gamma^A_1 e^{i \sum_{j=0}^{n-1} U_j}]
\end{equation}
For the Ising/Kitaev chain we may calculate this boundary unitary transform exactly:
\begin{equation}
{\mathcal U}_e= \cos \frac{\theta}{2}  + \cos \theta \sin \frac{\theta}{2}\,  \gamma^A_1 \sum_{j=2}^L
\left(\frac{h}{J}\right)^{j-2} \gamma^A_j
\label{eq:edgeU}
\end{equation}
where $\sin \theta = h/J$. 

Of course the ${\cal U}_e$ constructed in this manner is
not the full unitary transform shown to exist by the ADHH theorem, and the correspondingly transformed Hamiltonian would not display the emergent U(1) symmetry. Nevertheless this `boundary' unitary transform is sufficient to show conservation of the edge mode in the pre-thermal regime. Namely, if we combine the boundary unitaries from both ends, then the transformed Hamiltonian will conserve the $\mathbb{Z}_2$ bulk fermion parity $(\mathcal{U}_e\mathcal{U}'_e)^\dagger\prod_{j=1}^{L-1} \gamma_j^B \gamma_{j+1}^A\, \mathcal{U}_e\mathcal{U}'_e = (\mathcal{U}_e\mathcal{U}'_e)^\dagger \sigma^z_1 \sigma^z_L\, \mathcal{U}_e\mathcal{U}'_e $ up order $n_*=L$. As discussed in Section~\ref{sec:interacting_majorana}, this approximate conservation law is all that is needed.

Breaking the integrability by including non-zero $h_2$ and/or $J_2$ in (\ref{eqn:MZM-Y}), we  
have calculated ${\cal U}_e$ up to eleventh order using computer-aided algebra program. The resulting edge zero modes agree with those calculated explicitly in Ref.\ \onlinecite{Kemp17}. For example, when the only perturbing terms in $\hat{Y}$ are $h$ and $h_2$ we find to second order that:
\begin{align*}
\mathcal{U}_e=&\exp(\frac{1}{2}[ h \sigma^y_{1} \sigma^z_{2}+ h_2 \sigma^x_{1} \sigma^y_{2} \sigma^z_{3} + h^{2} \sigma^y_{1} \sigma^x_{2} \sigma^z_{3}\\ & +h h_2 \sigma^y_{1} (\sigma^z_{3} - \sigma^y_{2} \sigma^y_{3} \sigma^z_{4}) -h_2^2 \sigma^x_{1} \sigma^y_{2} \sigma^y_{3} \sigma^y_{4} \sigma^z_{5}]).
\end{align*}
Furthermore, this method of calculating the edge zero modes is preferable and more efficient than the method outlined in Ref.\ \onlinecite{Kemp17}, because the edge zero modes are automatically normalised at each order by construction.

It is also illuminating to implement the ADHH procedure directly and explicitly on the full bulk Hamiltonian. We show that for free-fermion Ising/Kitaev case the unitary transformation can be computed exactly. For simplicity, here we take $L\to\infty$ so the Hamiltonian of the Ising/Kitaev chain is
\begin{equation}
H_\text{IK} = -\sum_j\left(J \sigma^z_j \sigma^z_{j+1} +{h} \sigma^x_j\right)\ .
\label{HTFIM}
\end{equation}
We assume that $J$ is sufficiently large compared to $h$ so that we can apply ADHH with
\begin{equation}
\label{NIsingdef}
\hat{N}_{\rm IK} = \sum_{j} \sigma^z_i \sigma^z_{i+1}\ , \hskip 0.5 cm
\hat{Y}_{\rm IK} =  -{h}\sum_{j}\sigma^x_j\ .
\end{equation}
We rewrite the Hamiltonian in the form
\[H_{\rm IK} = -J\hat{N}_{\rm IK} -h \hat{D}_0 - i\frac{h}{2}\sum_j (\gamma^A_j\gamma^B_j+\gamma^B_{j-1}\gamma_{j+1}^A)
\]
so that the last term is the ``error'' term $\hat{E}$, which does not commute with $\hat{N}_{\rm IK}$
and
\[ \hat{D}_0 \equiv \frac{1}{2}\sum_j \sigma^x_j (1-\sigma^z_{j-1}\sigma^z_{j+1})=
\frac{i}{2}\sum_j  (\gamma^A_j\gamma^B_j-\gamma^B_{j-1}\gamma_{j+1}^A)
\]
commutes with $\hat{N}_{\rm IK}$. This can be checked explicitly; the key observation is that if there is
a single kink at either $j-1/2$ or $j+1/2$, flipping the spin at site $j$ hops the kink,
while with no kink or two kinks adjacent to $j$, flipping the spin creates or annihilates two kinks respectively.
The former process conserves $\hat{N}_{\rm IK}$, while the latter does not, so the former is allowed in $\hat{D}_{\rm IK}$.
Acting with the operator $(1-\sigma^z_{j-1}\sigma^z_{j+1})/2$ annihilates any configuration with zero or two kinks adjacent to $j$,
and gives the identity if there is a single one. Thus we arrive at the expression given above for $\hat{D}_0$.

At each step of the ADHH recursive procedure, new terms are included in $\hat{D}$ and the coefficient of the error term is reduced
by a power of $h/J$. At the very first recursive step, the error term can be canceled by using the relation
\[ [\hat{N}_{\rm IK} , \gamma^A_j\gamma^A_{j+1}]=-[\hat{N}_{\rm IK} , \gamma^B_{j-1}\gamma^B_{j}]=
2(\gamma^A_j\gamma^B_j+\gamma^B_{j-1}\gamma_{j+1}^A)\ .
\]
For simplicity, we impose periodic boundary conditions and define
\begin{align}
G_n=\frac{1}{4}\sum_j (\gamma^A_{j}\gamma^A_{j+n}- \gamma^B_j\gamma^B_{j+n})\ .
\label{Gdef}
\end{align}
Then, after the first step in the recursive procedure, the transformed Hamiltonian has the form given on the right-hand side of the
following equation:
\begin{align*}\left(1-\frac{h}{2J}G_1\right)&H_{\rm IK} \left(1+\frac{h}{2J}G_1\right)\\
&=  -J\hat{N}_{\rm IK} -h \hat{D}_0 + O\left(\frac{h^2}{J}\right)\ .
\end{align*}
Thus all terms in the transformed Hamiltonian that do not commute with $\hat{N}_{\rm IK}$ are at least of order 
$h^2/J$. The terms of this order can be split into pieces that do commute with $\hat{N}_{\rm IK}$, which then comprise $\hat{D}_1$, and those that do not, which comprise another error term. This procedure can be repeated, so that an order $(h/J)^2$ term can be added to ${\cal U}$ to yield an error term of order $h(h/J)^2$.

The ADHH theorem guarantees that this procedure can be repeated, at least up to order $n_*$.
Not surprisingly, the procedure can be implemented to all orders in a free-fermion system and so $n_*$ and hence $t_*$ become infinite as $L\to\infty$ when $h<J$  in the Ising/Kitaev chain.
This method was, in essence, how Onsager originally computed the free energy of the two-dimensional Ising model! His original calculations \cite{Onsager43} are manipulations of fermion bilinears, just as is required to find the unitary transformation ${\cal U}$. We find
\begin{align}
{\cal U} = \exp\left(-\sum_{n=1}^\infty \frac{1}{2n} \left(\frac{h}{J}\right)^n G_n\right)\ ,
\label{UIsing}
\end{align}
where the fermion bilinears $G_n$ are defined in (\ref{Gdef}).

The key to deriving (\ref{UIsing}) is to utilize the {\em Onsager algebra}, as derived in the original paper \cite{Onsager43}. This is simply the algebra of fermion bilinears, and a quick glance at the paper shows that Onsager defines them in terms of the conventional Jordan-Wigner expressions (given the clarity of his expression, one wonders why he missed defining individual fermions).  The generators of the algebra are then given by the $G_n$ in (\ref{Gdef}), and 
\begin{align}
A_m = -i\sum_j \gamma^B_{j}\gamma^A_{j+1-m}\ .
\label{Adef}
\end{align}
The Hamiltonian of the Ising/Kitaev chain $H_{\rm IK}=-J\hat{N}+\hat{Y}_{\rm IK}$ is then is written in terms of these generators as $\hat{N}_{\rm IK} = -A_0$ and $\hat{Y}_{\rm IK}=hA_1$. 
Onsager carefully works out the effect of periodic and antiperiodic boundary conditions, but we simplify matters here by taking $L\to\infty$.
It is then easy to work out the Onsager algebra
\begin{align}
\nonumber
[G_n,\, G_m]&=0\\
[G_n,\, A_m]&= A_{m+n}-A_{m-n}\\
\nonumber
[A_n,\, A_m]& = 8 G_{n-m}
\end{align}
for all integer $n$ and $m$.
These are exactly equations (61a), (61) and (60) of Ref.\ \onlinecite{Onsager43}, with a rescaling of the generators $G_n$. 

A key property of the Onsager algebra is that when $G_n$ is commuted with any linear combination of the $A_m$, the result remains linear in the $A_m$.  Moreover,
because by definition $G_{-n}=-G_{n}$, a series of quantities preserving the $U(1)$ symmetry is given by
$A_{m}+A_{-m}$.
This suggests then building the unitary
transformation out of the $G_n$, a task made even easier by the fact that they commute among themselves.
Such a construction is done straightforwardly by
Fourier-transforming the $A_m$ as
\begin{align} \tilde{A}(k) \equiv \sum_{m=-\infty}^\infty e^{-ikm} A_m\ ,
\end{align}
where $k$ takes values in the Brillouin zone $-\pi<k\le \pi$. Commuting any $G_n$ with $\tilde{A}(k)$ is then diagonal in $k$:
\begin{align}\left[G_n,\, \tilde{A}(k)\right] = (e^{-ink}-e^{ink})  \tilde{A}(k)\ . \end{align}
Because $G_n^\dagger=-G_n$, then
\[{\cal U}=e^{\sum_{n=1}^\infty u_n G_n}
\]
is indeed unitary when $u_n$ is real. Then
\begin{align}
{\cal U}\tilde{A}(k) {\cal U}^\dagger = \exp\left(\sum_{n=1}^\infty u_n(e^{-ink}-e^{ink})\right)\tilde{A}(k) \ .
\label{UAU}
\end{align}

Finding the appropriate coefficients $u_n$ for the ADHH transformation then becomes easy by rewriting the Hamiltonian using the inverse Fourier transformation as
\[
H_{\rm IK}=\int_{-\pi}^\pi \frac{dk}{2\pi} \big(J-he^{ik}\big) \tilde{A}(k)\ .
\]
Note then if we take $u_n =-(h/J)^n/(2n)$ as in (\ref{UIsing}), then (\ref{UAU}) gives
\[{\cal U}\tilde{A}(k) {\cal U}^\dagger =  \left(\frac{J-he^{-ik}}{J-he^{ik}}\right)^{1/2}
\tilde{A}(k)\ . \]
Thus, by making this choice, we find
\begin{align}
{\cal U}H_{\rm IK} {\cal U}^\dagger =  \int_{-\pi}^\pi \frac{dk}{2\pi} 
\left(J^2+h^2-2hJ\cos(k)\right)^{1/2} \tilde{A}(k) \ .
\label{UAU2}
\end{align}
This transformed Hamiltonian commutes with $\hat{N}_{\rm IK}$, since $\tilde{A}(k)+\tilde{A}(-k)$ does, and the function in the integrand of (\ref{UAU2}) is even in $k$. The unitary transformation in (\ref{UAU}) thus indeed does the job required of the ADHH theorem. It also shows that $n_*\to\infty$ for $L\to\infty$, since (\ref{UAU2}) shows that the error goes to zero in this limit.

In an integrable model such as the $XYZ$ model, there is no free-fermion representation, so the zero
mode can interact with bulk excitations. Nonetheless, it is possible to show by rewriting the exact strong zero mode of Ref.\ \onlinecite{Fendley16} as a matrix-product operator that $n_*$ is indeed infinite here as well \cite{Verstraete17}, at least for the edge mode. That tantalizingly suggests that $n_*$ is infinite for all integrable systems.

\section{Discussion}
We showed that prethermalization can extend
topological protection into regimes where it might have been expected to fail.
Perhaps the simplest experimental realization of the prethermal protection of edge modes,
albeit in a system without topological order, is in a trapped ion or neutral atom chain governed by
a perturbed transverse-field Ising model. In a solid state system, prethermalization
is ultimately interrupted by the thermalization driven by the electron-phonon interaction.
However, over intermediate time scales
a chain of quantum dots tuned to a Kitaev-type Hamiltonian may exhibit
prethermal strong zero modes that survive until the late time at which
the electron-phonon interaction causes thermalization.
We emphasize that prethermalization can occur in any dimension, and the long-lived zero modes can,
as a result, occur in two or three dimensions. Our work is therefore relevant to the proposal of Ref. \onlinecite{Barkeshli15} for a universal topological quantum computer. 

In a more theoretical direction, we note that 
the ADHH theorem applies to much more general systems than those with topological order. It provides a precise and rigorous approach to finding an approximately conserved charge. Integrable models are so because of the presence of non-trivial conserved charges, and the resemblance of the Onsager analysis of the Ising model to the results derived here hints at a more general connection between the ADHH procedure and integrability. For example, the $U(1)$ quantum number conservation implied by the theorem is strongly reminiscent of the conservation of quasiparticle number in integrable field theories. Moreover, the general connection between integrability breaking and prethermalization suggests that the extremely long lifetime of the prethermal strong zero modes is a consequence of the structure of integrability still affecting the perturbed system, at least in one dimension. In higher dimensions, the role of integrability seemingly is being played by the requirement of integer eigenvalues of $\hat{N}$, but why this is so is somewhat mysterious. Clearly more research in both more formal and experimental directions is warranted.

\begin{acknowledgments}
We would like to thank A. Akhmerov, S. Frolov, P. Hess, K. van Hoogdalem, L. Kouwenhoven, and F.\ Verstraete for valuable discussions. D. Else was supported by the Microsoft Corporation, while the work of P.F.\ was supported by EPSRC through grant EP/N01930X.

\end{acknowledgments}

\appendix
\setcounter{equation}{0}
\renewcommand{\theequation}{\Alph{section}\arabic{equation}}

\section{Extending the ADHH theorem to non-single-site $\hat{N}$}
\label{app:nonsinglesite}
Here we will explain how the proof of ADHH can extended to the case where the unperturbed Hamiltonian $\hat{N}$ can be written as a sum $\hat{N}_\Gamma = \sum_\Gamma \hat{N}_\Gamma$, where each $\hat{N}_\Gamma$ is supported on a set of sites $\Gamma$ of finite radius. Recall that the original proof of ADHH assumed that each $\hat{N}_\Gamma$ acts on just a single site. The only place where this assumption was used is Section 5.4 of Ref.~\onlinecite{Abanin15b}, in which it was implicitly assumed that
\begin{equation}
\label{an_equation}
\| e^{is\hat{N}} V e^{-is\hat{N}} \|_{\kappa} = \| V \|_\kappa,
\end{equation}
where $\| \cdot \|_\kappa$ is the norm for potentials introduced in Section 2.2 of Ref.~\onlinecite{Abanin15b}.
This assumption no longer holds if the terms of $\hat{N}$ do not act on one site each, because then evolution by $\hat{N}$ does not preserve the support of local terms.

We can get around this problem quite straightforwardly. We say that an operator $V_Z$ is ``strongly supported'' on a set $Z$ if $V_Z$ is supported on the set $Z$, \emph{and} $V_Z$ also commutes with all $\hat{N}_{\Gamma}$'s such that $\Gamma \nsubseteq Z$. For $\hat{N}$ finite range, this just requires increasing the size of $Z$ slightly. We then redefine the norm $\|\cdot\|_\kappa$ so that it is based on the strong support of local terms rather than the support. With this modification, we see that \eqnref{an_equation} is recovered, since the ``strong support'' of an operator does not grow under $\hat{N}$. We observe that if $A_Z$ and $B_{Z'}$ are strongly supported on sets $Z$, $Z'$, then $[A_Z, B_{Z'}] = 0$ if $Z \cap Z' = \emptyset$, and moreover in general $[A_Z, B_{Z'}]$ is strongly supported on $Z \cup Z'$. This allows the rest of the proof in Ref.~\onlinecite{Abanin15b} to carry over without change.

\section{Energy bandwidth}
\label{appendix:energy_bandwidth}
Here we prove the claim we made about the energy bandwidth of the $\hat{M} = m$ sector under the Hamiltonian $J\hat{M} + \hat{D}$. The starting point is the observation that a consequence of Ref. \onlinecite{Abanin15b} is a bound on the ``local norm'' of $\hat{D}$. Namely,
\begin{equation}
  \label{eqn:Dbound}
  \| \hat{D} \|_{0} \leq C J_0\ ,
\end{equation}
where we define
$J_0 \equiv \frac{1}{\kappa_0^2} \| \hat{Y} \|_{\kappa_0}$
and fix $\kappa_0$ such that $J_0 < \infty$. The dimensionless constant $C$ is proportional to $\kappa_0^2$.
We define the local norm $\|\mathcal{H}\|_\kappa$ of a Hamiltonian $\mathcal{H} = \sum_\Gamma \mathcal{H}_\Gamma$ (where the $\Gamma$ are subsets of the lattice $\Lambda$, and $\mathcal{H}_\Gamma$ is an operator supported on $\Gamma$), as
\begin{equation}
\| \mathcal{H} \|_{\kappa} = \sup_{x \in \Lambda} \sum_{\Gamma \ni x} e^{\kappa |\Gamma|} \| \mathcal{H}_\Gamma \|.
\end{equation}
We derived Eq.~(\ref{eqn:Dbound}) by applying the bound in the unnumbered equation just after Eqn. (4.10)
of Ref.~\onlinecite{Abanin15b}, invoked the fact
that $\| \cdot \|_0 \leq \| \cdot \|_\kappa$ for $\kappa > 0$, and then summed over
$n$.

We will state our results in some degree of generality. Rather than considering any particular form of $\hat{M}$, we just assume that it can be written as
\begin{equation}
\hat{M} = \sum_{x \in \mathcal{X}} P_x,
\end{equation}
where the $P_x$'s are commuting projectors, and $\mathcal{X}$ is some set to index the projectors. We assume that each projector $P_x$ is supported on a set $B_x \subseteq \Lambda$ (where $\Lambda$ is the set of all sites in the lattice).

We can label eigenstates of $\hat{M}$ by their simultaneous eigenvalue under the projectors $P_x$, which we refer to as a ``syndrome''. More precisely, a syndrome is a subset $S \subseteq \mathcal{X}$ whose corresponding projectors are not satisfied.
Moreover, for each syndrome $s$ we can construct a corresponding projector $P_s \equiv \left(\prod_{x \in S} P_x\right) \left(\prod_{x \notin s} (1-P_x)\right)$. We let $P$ denote the projector corresponding to the trivial syndrome, $P \equiv P_\emptyset$ (i.e. the projector onto the ground state subspace of $\hat{N}$).

We also introduce the notion of a \emph{partial syndrome} $(\mathcal{Y}, s_\mathcal{Y})$ where $\mathcal{Y} \subseteq \mathcal{X}$ and $s_\mathcal{Y} \subseteq \mathcal{Y}$. We say that $(\mathcal{Y}, s_\mathcal{Y})$ is the \emph{restriction} of a syndrome $s$ if $s_\mathcal{Y} = \mathcal{Y} \cap s$. A partial syndrome specifies the eigenvalue of only those projectors indexed by $x \in \mathcal{Y}$.
 The projector corresponding to a partial syndrome is
\begin{equation}
Q_{\mathcal{Y}, s_{\mathcal{Y}}} = \left(\prod_{x \in s_{\mathcal{Y}}} P_x \right) \left(\prod_{x \in \mathcal{Y} \setminus s_{\mathcal{Y}}} (1-P_x) \right)
\end{equation}

We now state the following condition under which we will prove our results.

\begin{framed}
\textbf{Local-TQO}. There exists a constant $K$ such that, for any syndrome $s$, there exists a region $\mathcal{R}_s$, of size at most $K|s|$, such that $B_x \subseteq \mathcal{R}_s$ for all $x \in s$, and furthermore for any region $\Gamma$ with $\Gamma \cap \mathcal{R}_s = \emptyset$, we have
\begin{equation}
\label{localtqo}
Q_{\mathcal{A},s_\mathcal{A}} 
V_\Gamma
Q_{\mathcal{A},s_{\mathcal{A}}}
= c(V_\Gamma) Q_{\mathcal{A},s_{\mathcal{A}}}\,
\end{equation}
where $c(V_\Gamma) = \Tr(P V_\Gamma P)/\Tr(P)$, and 
$\mathcal{A} = \mathcal{X} \setminus s$, which implies that the restriction $s_{\mathcal{A}} = s \cap \mathcal{A} = \emptyset$.
\end{framed}

Roughly, this is saying that there is a topologically protected degeneracy in
the ground-state of $\hat{M}$ (by applying it to the trivial syndrome), and
moreover that an excited state with eigenvalue $m$
is localized to a region of size at most $Km$, and looks like the ground state elsewhere. The condition ``Local-TQO'' is closely related to the conditions under which stability of the topological order in the ground state subspace of $\hat{M}$ was proven in Refs.~\onlinecite{Bravyi10,Bravyi11,Michalakis13} (though the result we want to prove here is somewhat different). We observe that ``Local-TQO'' is indeed satisfied for the $\hat{N}$ of the Kitaev chain [\eqnref{eqn:MZM-L}], with $K=2$.

Now we can prove the following theorem:
\begin{thm}
If the condition ``Local-TQO'' is satisfied, then for any Hamiltonian $V =
  \sum_\Gamma V_\Gamma$ that commutes with $\hat{M}$, the spectrum of
  $J\hat{M}+V$ in the eigenspace of $\hat{M}$ with eigenvalue $m$ lies within the interval
\begin{equation}
\biggl[c(V) + m(J-K\|V\|_0) \; , \; c(V) + m(J+ K\|V\|_0)\biggr].
\end{equation}

\begin{proof}
Consider an operator $V_\Gamma$ supported on a set $\Gamma \subseteq \Lambda$. We want to consider the circumstances under which $P_s V_\Gamma P_{t} - \delta_{s,t} c(V_\Gamma) P_s$ can be nonzero for two syndromes $s,t \in \mathcal{S}$ such that $|s| = |t|$.
Let us partition the set $\mathcal{X}$ into $\mathcal{X}_\Gamma$ and $\mathcal{X}_\Gamma^c$, where $\mathcal{X}_\Gamma = \{ x \in \mathcal{X} : B_x \cap \Gamma \neq \emptyset \}$. 
 We consider the following cases:
\begin{itemize}
\item \emph{$s \neq t$ and $s \cap \mathcal{X}_\Gamma^c \neq t \cap \mathcal{X}_\Gamma^c$}.

Without loss of generality, we can say that there exists $x \in s \cap \mathcal{X}_\Gamma^c$ such that $x \notin t$. Then we note that $P_s P_x = 1$ and $P_x P_t = 0$. Furthermore, $x \in \mathcal{X}_\Gamma^c$ implies that $[P_x, V_\Gamma] = 0$. Hence, we can write $P_s V_\Gamma P_t = P_s P_x V_\Gamma P_t = P_s V_\Gamma P_x P_t = 0$.
\item $s \neq t$, $s \cap \mathcal{X}_\Gamma^c = t \cap \mathcal{X}_\Gamma^c$ and $\mathcal{R}_s \cap \Gamma = \emptyset$.

$\mathcal{R}_s \cap \Gamma = \emptyset$ implies that for all $x \in s$, $B_x \cap \Gamma = \emptyset$ (since $B_x \subseteq R_s$). This implies that $s \subseteq \mathcal{X}_\Gamma^c$, and hence that $s = s \cap \mathcal{X}_\Gamma^c = t \cap \mathcal{X}_\Gamma^c$. Since $|s| = |t|$ this implies that $s \neq t$ which contradicts our assumption.

\item $s = t$ and $\mathcal{R}_s \cap \Gamma = \emptyset$.

 We decompose $s$ into partial syndromes $(\mathcal{A}, s_{\mathcal{A}})$ and $(\mathcal{B}, s_{\mathcal{B}})$ 
 where $\mathcal{A} = s$
and $\mathcal{B} = \mathcal{X} \setminus \mathcal{A}$. Then we can write $P_s = Q_{\mathcal{A}, s_{\mathcal{A}}} Q_{\mathcal{B}, s_{\mathcal{B}}}$. We observe that $\mathcal{R}_s \cap \Gamma \neq \emptyset$ ensures that $Q_{\mathcal{B}, s_{\mathcal{B}}}$ commutes with $V_\Gamma$. Hence, we find that
\begin{align}
\nonumber
P_s V_\Gamma P_s &= Q_{\mathcal{B}, s_{\mathcal{B}}} (Q_{\mathcal{A}, s_{\mathcal{A}}} V_\Gamma Q_{\mathcal{A}, s_{\mathcal{A}}}) Q_{\mathcal{B}, s_{\mathcal{B}}} \\
&= c(V_\Gamma) P_s,
\end{align}
by \eqnref{localtqo}.
\end{itemize}
In conclusion, we find that $P_s V_\Gamma P_t - \delta_{s,t} c(V_\Gamma) P_s = 0$ \emph{except} when $\mathcal{R}_s \cap \Gamma \neq \emptyset$ and $\mathcal{R}_t\cap \Gamma \neq \emptyset$.

Now consider a Hamiltonian $V = \sum_\Gamma V_\Gamma$. Let $\mathbb{P}_m = \sum_{s \in \mathcal{S} : |s| = E} P_s$. (That is, $\mathbb{P}_m$ is the projector onto the subspace with eigenvalue $m$ under $\hat{M}$). Furthermore, let $\mathbb{P}_m^{\Gamma} = \sum_{s \in \mathcal{S}, |s| = m, \mathcal{R}_s \cap \Gamma \neq \emptyset} P_m$. Then we see that
\begin{align*}
\mathbb{P}_m V \mathbb{P}_m - c(V) \mathbb{P}_m &= 
\sum_\Gamma \mathbb{P}_m V_\Gamma \mathbb{P}_m - c(V_\Gamma) \mathbb{P}_n \\
&= \sum_\Gamma \mathbb{P}_m^{\Gamma} V_\Gamma \mathbb{P}_m^{\Gamma} - c(V_\Gamma) \mathbb{P}_{m}^{\Gamma} \\
&\leq \sum_\Gamma \| V_\Gamma \| \mathbb{P}_m^{\Gamma}\\ 
&= \sum_{s : |s| = m} \left(\sum_{\Gamma: \mathcal{R}_s \cap \Gamma \neq 0} \| V_\Gamma \| \right) P_s \\
&\leq \sum_{s : |s| = m} |\mathcal{R}_s| \|V\|_0 P_s \\
&\leq Km \|V\|_0 \sum_{s : |s| = m} P_s \\
&= Km\|V\|_0 \mathbb{P}_m
\end{align*}
Hence,
\begin{align}
\| \mathbb{P}_m V \mathbb{P}_m - c(V) \mathbb{P}_m\| \leq Km \|V\|_0 \ .
\end{align}
The theorem immediately follows.
\end{proof}
\end{thm}

\bibliography{topo-phases}

\clearpage

\end{document}